\pdfoutput=1
\documentclass[twocolumn]{aastex701}
\usepackage{amsmath}
\usepackage{tikz}
\usetikzlibrary{arrows.meta}
\usepackage{listings}
\usepackage{xcolor}
\newcommand{\rmax}{r_{\mathrm{max}}}
\newcommand{\Mpch}{\,h^{-1}\mathrm{Mpc}}

\shorttitle{Fast Graph-based $N$-point statistics on the GPU}
\shortauthors{C.~G.~Sabiu }

\begin{document}

\title{Fast Graph-based Higher-Order Clustering Statistics on the GPU}

\author[0000-0002-5513-5303]{Cristiano G.\ Sabiu}
\affiliation{Natural Science Research Institute (NSRI), University of Seoul, Seoul 02504, Korea}
\email{csabiu@gmail.com}

\begin{abstract}
We present a significant update to \texttt{GRAMSCI} \citep[GRAph Made Statistics for Cosmological
Information; ][]{sabiu19}, an algorithm for the fast computation of the general $N$-point spatial correlation function of any
discrete point set.
Version 2 adds several features.
We replace the binary-search inner loop
($O(m\log m)$ per hub--spoke pair, $m$ being the mean neighbor count) with
a {\em merge-walk} algorithm that reduces the inner loop to $O(m)$. We
implement a parity-decomposed 4pCF that separates the signal into even
and odd channels. We estimate the disconnected 4pCF internally on the same graph to return the connected 4pCF.
We add a Python/NumPy interface to the Fortran engine. Principally, we
present a GPU port of the full query engine (OpenACC): the 3pCF, 4pCF, and
parity-decomposed 4pCF kernels run on a single
consumer GPU with measured speedups of $2.6\times$ (3pCF) to $9\times$
(4pCF) over a 64-thread CPU node, and out-of-core tiling handles graphs exceeding device memory. We measure a $9\times10^9$-edge
BAO-scale 3pCF on a 24\,GB card with ${\sim}20\%$ overhead. We validate the
code against its CPU reference, against analytic injection tests,
and on exactly solvable random fields (a non-Gaussian
multiplicative cascade and a Gaussian field, whose full and connected 4pCF
the estimators reproduce without bias), and demonstrate BAO-scale applications on the
DESI DR1 LRG sample (the 3pCF and the full and connected 4pCF over the complete configuration space, both to $\rmax = 150\Mpch$).
\end{abstract}


\section{Introduction}
\label{sec:intro}

The spatial distribution of galaxies encodes a wealth of information about
the primordial density field and its subsequent gravitational evolution.
The two-point correlation function (2pCF), which measures the excess
probability of finding galaxy pairs at a given separation over a random
distribution, has long been the workhorse of this programme. Early analyses
of the angular and spatial clustering of galaxies established that the 2pCF
is well described by a power law, $\xi(r) = (r/r_0)^{-\gamma}$, with a
correlation length $r_0 \sim 5\Mpch$ and slope $\gamma \simeq 1.8$
\citep{totsuji69,groth77,peebles75,davis83}. The practical measurement of
$\xi(r)$ from a flux-limited catalogue relies on pair-counting estimators
that correct for the survey geometry and edge effects, of which the
\citet{landy93} estimator and its variants \citep{hamilton93,szapudi98} are
now standard. Beyond characterizing galaxy clustering and its dependence on
luminosity and color through halo occupation distribution modeling
\citep{zehavi11}, the 2pCF underpins much of modern survey cosmology. A
landmark example is the detection of the baryon acoustic oscillation (BAO)
peak at ${\sim}100\Mpch$ in the configuration-space 2pCF of SDSS luminous
red galaxies \citep{eisenstein05}, which provides a standard ruler for the
expansion history and has since become a cornerstone of large galaxy-survey
analyses \citep[e.g.][]{anderson14,alam17}.

Powerful as it is, the 2pCF does not exhaust the information content of the
galaxy field. For a Gaussian random field the 2pCF, together with its
Fourier counterpart the power spectrum, provides a complete statistical
description, and all higher-order correlations either vanish or are fixed by
the two-point function. The late-time galaxy distribution, however, is
manifestly non-Gaussian. Nonlinear gravitational evolution couples Fourier
modes and generates connected higher-order moments even from Gaussian
initial conditions, while the biased and nonlinear relation between galaxies
and the underlying matter field imprints further non-Gaussian structure
\citep{fry93}. Higher-order statistics therefore carry complementary
information that is, by construction, inaccessible to the power spectrum
alone.

The lowest-order such probe is the three-point correlation function (3pCF),
the connected triplet statistic whose Fourier counterpart is the bispectrum,
first measured in galaxy catalogues by \citet{groth77}. Because
gravitational and bias contributions enter the 3pCF with different
configuration dependence, it breaks the degeneracy between the clustering
amplitude and the galaxy bias that limits 2pCF-only analyses
\citep{frieman94,gaztanaga05}, and so provides a direct handle on the
galaxy-matter connection. This has been exploited to measure the linear and
nonlinear bias parameters from redshift surveys \citep{verde02,guo15}, to
constrain primordial non-Gaussianity \citep{sefusatti06}, and, more
recently, to track the BAO feature at third order
\citep{gaztanaga09,slepian17,slepian17a,gilmarin17}, adding leverage on the
standard ruler that is independent of the 2pCF. The connected four-point
correlation function (4pCF) is related to the primordial trispectrum and
provides additional leverage on inflationary models \citep{sefusatti07}.
Using the Quijote simulations, \citet{kim26} quantified the cosmological
information that the configuration-space 3pCF and connected 4pCF of dark
matter halos add to the 2pCF.

Naively, computing $N$-point functions from large catalogues is expensive.
A brute-force approach requires $O(N^k)$ operations for a $k$-point
function. Thankfully, tree-based methods reduce this to $O(Nm^{k-1})$, where
$m$ is the mean number of neighbors within the maximum scale $\rmax$,
although the $m^{k-1}$ factor still grows rapidly.

Several codes are publicly available. \texttt{TreeCorr} \citep{jarvis04}
provides optimized 2pCF and 3pCF estimators for projected and
three-dimensional catalogues. \texttt{Corrfunc} \citep{sinha20} uses SIMD
vectorization (AVX2/AVX-512) for maximum 2pCF throughput.
\texttt{ConKer} \citep{brown22} estimates isotropic
correlation functions of arbitrary order by FFT convolution of the
gridded density field with spherical kernels, achieving favorable
scaling for angle-averaged statistics. A complementary
algorithmic line, initiated by \citet{slepian15} and accelerated with FFTs
by \citet{slepian16}, expands the correlators in a spherical-harmonic
basis. \texttt{Encore} \citep{philcox22} extends this approach to the
$N$-pCF, replacing direct enumeration with matrix products. This achieves
favorable scaling at high neighbor density but does not directly yield a
configuration-space connected 4pCF with parity decomposition.

\texttt{GRAMSCI} \citep{sabiu19} uses a graph-database approach: it builds a
compact
sorted adjacency structure once from a kd-tree, then queries it for any
combination of 2-, 3-, and 4-point statistics without revisiting the
original coordinates. The outline of this paper is as follows. In
\S\ref{sec:algorithm} we briefly summarize the \texttt{GRAMSCI} algorithm. In
\S\ref{sec:new} we detail the four directions in which this paper extends
the original work: an $O(m)$ merge-walk inner kernel (\S\ref{sec:mergewalk}),
a parity-decomposed 4pCF (\S\ref{sec:parity}), a connected 4pCF with internal disconnected
subtraction (\S\ref{sec:connected}), a Python interface
(\S\ref{sec:python}), and a GPU port of the complete query engine with
out-of-core support for graphs exceeding device memory (\S\ref{sec:gpu}). In
\S\ref{sec:performance} we benchmark the new code for speed, scalability and
correctness. We then apply our algorithm to observational and simulated data; the DESI DR1 LRG sample and mock galaxy catalogues (\S\ref{sec:desi}). 
We discuss the scientific applications in \S\ref{sec:applications} and then conclude in
\S\ref{sec:summary}.

\section{The GRAMSCI Algorithm}
\label{sec:algorithm}

In an effort to make this paper self-contained we summarize the algorithm
here; full details are in \citet{sabiu19}.

\subsection{Graph construction}
\label{sec:graph}

Given a catalogue of $N$ objects, \texttt{GRAMSCI} builds a kd-tree and finds,
for
each object $i$, all neighbors within $\rmax$. These are stored as a sorted
adjacency list:
\begin{equation}
  \mathcal{N}(i) = \{j_1, j_2, \ldots, j_{m_i}\}, \quad
  j_1 < j_2 < \cdots < j_{m_i}.
  \label{eq:adjacency}
\end{equation}
Alongside each neighbor index, the radial-bin index (an integer byte) and
optionally a two-byte direction-pixel index (for the parity 4pCF) are stored. Counts
are accumulated with signed, normalized data-minus-random weights in the
style of the \citet{landy93} and \citet{szapudi98} estimators.
Section~\ref{sec:window} defines the resulting
estimator explicitly and discusses its normalization, including the
parity-decomposed case, and its behavior under a realistic survey
window. The kd-tree
and raw coordinates are then discarded; all subsequent NpCF queries operate
entirely on the adjacency lists.

The total memory footprint is $O(Nm)$ rather than $O(N^2)$. The sorted order
in Equation~(\ref{eq:adjacency}) is a key invariant exploited by the
merge-walk kernel and the GPU binary-search kernels alike.

\subsection{Triangle and tetrahedron enumeration}
\label{sec:enumeration}

For the 3pCF, \texttt{GRAMSCI} uses a hub-and-spoke loop: for each hub $i$ and
each
neighbor $j \in \mathcal{N}(i)$, it checks whether a second neighbor
$k \in \mathcal{N}(i)$ with $k > j$ is also in $\mathcal{N}(j)$. If so, the
triangle $(i,j,k)$ is counted in the appropriate $(r_{ij}, r_{ik}, r_{jk})$
bin. The 4pCF extends this to tetrahedra $(i, j_1, j_2, j_3)$ where every
pair is connected. The original membership test was a binary search on
$\mathcal{N}(j)$, costing $O(\log m)$ per candidate.
Figure~\ref{fig:mergewalk}a illustrates the hub--spoke
structure and the role of the sorted adjacency lists.

\section{New Contributions}
\label{sec:new}

\subsection{Merge-walk: \texorpdfstring{$O(m)$}{O(m)} sorted-set intersection}
\label{sec:mergewalk}

The original membership test ran a binary search for every candidate.
Thankfully the sorted adjacency lists let us replace it with something
cheaper. Since the hub tail $\{k \in \mathcal{N}(i) : k > j\}$ and
$\mathcal{N}(j)$
are both sorted by node index (Equation~\ref{eq:adjacency}), their
intersection can be found in a single linear pass with two pointers $a$
and $b$:
\begin{itemize}
\item $a$ walks the hub tail, starting at the entry after $j$.
\item $b$ walks $\mathcal{N}(j)$ from its beginning.
\end{itemize}
At each step:
\begin{itemize}
\item $\mathcal{N}(i)[a] = \mathcal{N}(j)[b]$: triangle found; accumulate
  the $(r_{ij}, r_{ik}, r_{jk})$ bin contribution; advance both.
\item $\mathcal{N}(i)[a] < \mathcal{N}(j)[b]$: advance $a$.
\item otherwise: advance $b$.
\end{itemize}
Each list is visited at most once, giving $O(m_i + m_j)$ cost per spoke,
an $O(\log m)$ improvement over binary search
(Figure~\ref{fig:mergewalk}b).

\subsubsection{Three-way merge for the 4pCF}
\label{sec:threeway}

For the 4pCF, once the k2 walk has located a second spoke $j_2$, the k3
loop must find a third spoke $j_3$ that appears simultaneously in the hub
tail beyond $j_2$, in $\mathcal{N}(j_1)$, and in $\mathcal{N}(j_2)$. We
generalize the two-pointer walk to three pointers $\alpha, \beta, \gamma$:
\begin{itemize}
\item $\alpha$: hub tail beyond $j_2$.
\item $\beta$: $\mathcal{N}(j_1)$, starting at the position just after
  $j_2$ (reused from the k2 walk, so no rescanning).
\item $\gamma$: $\mathcal{N}(j_2)$, pre-advanced past $j_2$'s own index
  (entries $\le j_2$ can never match the hub tail).
\end{itemize}
Let $h_\alpha, h_\beta, h_\gamma$ be the current candidate node indices.
If all three are equal, we record the tetrahedron and advance all three.
Otherwise every pointer at the current minimum is advanced:
\begin{equation}
\begin{aligned}
  h_\alpha \le h_\beta \text{ and } h_\alpha \le h_\gamma &: \quad
    \alpha \leftarrow \alpha + 1 \\
  h_\beta \le h_\alpha \text{ and } h_\beta \le h_\gamma &: \quad
    \beta \leftarrow \beta + 1 \\
  h_\gamma \le h_\alpha \text{ and } h_\gamma \le h_\beta &: \quad
    \gamma \leftarrow \gamma + 1
\end{aligned}
\label{eq:threeway}
\end{equation}
(Multiple conditions can be true simultaneously when two minima are tied;
in that case multiple pointers are advanced in the same step.) This visits
each list at most once, giving $O(m)$ cost per $(i, j_1, j_2)$ triple
compared with two $O(\log m)$ binary searches in the original code.
Figure~\ref{fig:mergewalk} illustrates the complete
scheme (hub, spokes, sorted adjacency lists, and both walks) on a
small worked example.

\begin{figure*}[t!]
\centering
\begin{tikzpicture}[scale=1.12,
  gal/.style={circle, fill=black, inner sep=1.4pt},
  hub/.style={circle, fill=red!80!black, inner sep=1.8pt},
  lbl/.style={font=\scriptsize}]
  \draw[dashed, gray] (0,0) circle (1.6);
  \node[gray, font=\scriptsize] at (1.25, 1.28) {$\rmax$};
  \fill[blue!9] (0,0) -- (-0.5,1.05) -- (0.62,0.55) -- cycle;
  \draw[very thick, blue!60!black]
    (0,0) -- (-0.5,1.05)  (0,0) -- (0.62,0.55)  (0,0) -- (0.35,1.42)
    (-0.5,1.05) -- (0.62,0.55)  (-0.5,1.05) -- (0.35,1.42)
    (0.62,0.55) -- (0.35,1.42);
  \draw[semithick] (0,0) -- (-1.05,-0.85)  (0,0) -- (1.4,-0.3);
  \draw[thin, gray] (0.62,0.55) -- (1.4,-0.3);
  \draw[thin, gray] (-0.5,1.05) -- (-1.55,1.55);
  \node[hub, label={[lbl]below:{$2$ (hub $i$)}}]  at (0,0) {};
  \node[gal, label={[lbl]left:{$5$ ($j_1$)}}]     at (-0.5,1.05) {};
  \node[gal, label={[lbl]right:{$9$ ($j_2$)}}]    at (0.62,0.55) {};
  \node[gal, label={[lbl]above:{$14$ ($j_3$)}}]   at (0.35,1.42) {};
  \node[gal, label={[lbl]below:{$11$}}]           at (-1.05,-0.85) {};
  \node[gal, label={[lbl]right:{$17$}}]           at (1.4,-0.3) {};
  \node[gal, label={[lbl]above:{$21$}}]           at (-1.55,1.55) {};
  \node[font=\small] at (-1.5,-1.7) {(a)};
\end{tikzpicture}
\hfill
\begin{tikzpicture}[scale=1.0,
  cell/.style={draw, minimum width=0.5cm, minimum height=0.42cm,
               inner sep=1pt, font=\scriptsize},
  match/.style={cell, fill=green!25},
  dim/.style={cell, fill=black!8, text=black!45},
  lbl/.style={font=\scriptsize}]
  \node[lbl, anchor=east] at (-0.15, 1.0) {$\mathcal{N}(2)_{>j_1}$:};
  \node[match] (h1) at (0.25,1.0) {9};
  \node[cell]  (h2) at (0.75,1.0) {11};
  \node[match] (h3) at (1.25,1.0) {14};
  \node[cell]  (h4) at (1.75,1.0) {17};
  \node[lbl, anchor=east] at (-0.15, 0.0) {$\mathcal{N}(5)$:};
  \node[dim]   (n1) at (0.25,0.0) {2};
  \node[match] (n2) at (0.75,0.0) {9};
  \node[match] (n3) at (1.25,0.0) {14};
  \node[cell]  (n4) at (1.75,0.0) {21};
  \draw[green!55!black, semithick] (h1.south) -- (n2.north);
  \draw[green!55!black, semithick] (h3.south) -- (n3.north);
  \draw[-stealth, thick] (0.25,1.62) node[above, lbl] {$a$} -- (h1.north);
  \draw[-stealth, thick] (0.25,-0.62) node[below, lbl] {$b$} -- (n1.south);
  \draw[-stealth, dashed, black!60] (0.5,1.85) -- (1.75,1.85)
    node[midway, above, lbl] {advance};
  \draw[-stealth, dashed, black!60] (0.5,-0.9) -- (1.75,-0.9);
  \node[lbl, align=left, anchor=north west] at (-0.9,-1.25)
    {matches $\Rightarrow$ triangles $(2,5,9)$, $(2,5,14)$};
  \node[font=\small] at (-0.75,-2.0) {(b)};
\end{tikzpicture}
\hfill
\begin{tikzpicture}[scale=1.0,
  cell/.style={draw, minimum width=0.5cm, minimum height=0.42cm,
               inner sep=1pt, font=\scriptsize},
  match/.style={cell, fill=green!25},
  lbl/.style={font=\scriptsize}]
  \node[lbl, anchor=east] at (-0.15, 1.6) {$\mathcal{N}(2)_{>j_2}$:};
  \node[cell]  (a1) at (0.25,1.6) {11};
  \node[match] (a2) at (0.75,1.6) {14};
  \node[cell]  (a3) at (1.25,1.6) {17};
  \node[lbl, anchor=east] at (-0.15, 0.8) {$\mathcal{N}(5)_{>j_2}$:};
  \node[match] (b1) at (0.25,0.8) {14};
  \node[cell]  (b2) at (0.75,0.8) {21};
  \node[lbl, anchor=east] at (-0.15, 0.0) {$\mathcal{N}(9)_{>j_2}$:};
  \node[match] (c1) at (0.25,0.0) {14};
  \node[cell]  (c2) at (0.75,0.0) {17};
  \draw[green!55!black, semithick] (a2.south) -- (b1.north);
  \draw[green!55!black, semithick] (b1.south) -- (c1.north);
  \draw[-stealth, thick] (0.25,2.22) node[above, lbl] {$\alpha$} -- (a1.north);
  \draw[-stealth, thick] (-0.6,0.33) node[left, lbl] {$\beta$}
     -- (b1.south west);
  \draw[-stealth, thick] (0.25,-0.62) node[below, lbl] {$\gamma$} -- (c1.south);
  \node[lbl, align=left, anchor=north west] at (-0.9,-1.25)
    {triple match $\Rightarrow$ tetrahedron $(2,5,9,14)$};
  \node[font=\small] at (-0.75,-2.0) {(c)};
\end{tikzpicture}
\caption{The graph-based enumeration on a worked example.
\textit{(a)} The hub $i = 2$ (red) and its sorted adjacency list
$\mathcal{N}(2) = \{5, 9, 11, 14, 17\}$, all points within $\rmax$
(dashed), with the spokes drawn to each neighbor.  Point 21 lies
outside the hub's ball but inside $\mathcal{N}(5)$; grey edges belong to
the graph but not to any tuple rooted at this hub--spoke choice.
\textit{(b)} The two-pointer merge walk for spoke $j_1 = 5$: pointer $a$
walks the hub tail $\mathcal{N}(2)_{>j_1} = \{9, 11, 14, 17\}$ and $b$
walks $\mathcal{N}(5) = \{2, 9, 14, 21\}$; at each step the pointer at
the smaller index advances, and equal indices (green) are triangles:
here $(2,5,9)$ and $(2,5,14)$, the shaded triangle and part of the bold
tetrahedron in panel (a).  Entries below the hub tail's range (grey) are
skipped by the same rule, never searched.  \textit{(c)} Once the walk in
(b) has located $j_2 = 9$, the three-pointer walk of
Equation~(\ref{eq:threeway}) runs $\alpha$ over the hub tail beyond
$j_2$, $\beta$ over $\mathcal{N}(5)$ beyond $j_2$ (position reused from
the $k_2$ walk), and $\gamma$ over $\mathcal{N}(9)$ past its own index
(Section~\ref{sec:threeway}); a triple match (green) is a
tetrahedron: here $(2,5,9,14)$, bold in panel (a).  Each list is
visited at most once per walk, giving the $O(m)$ inner-loop cost.
\label{fig:mergewalk}}
\end{figure*}

\subsection{Parity-decomposed 4pCF}
\label{sec:parity}

A standard 4pCF estimator gives one number per tetrahedron configuration.
We add a parity decomposition that separates the signal into
parity-even ($\zeta^{+}$) and parity-odd ($\zeta^{-}$) channels. For each
tetrahedron $(i, j_1, j_2, j_3)$, the handedness is determined by the
signed volume
\begin{equation}
  V = (\mathbf{r}_{ij_1} \times \mathbf{r}_{ij_2}) \cdot \mathbf{r}_{ij_3}.
  \label{eq:volume}
\end{equation}
Figure~\ref{fig:chirality} illustrates the underlying geometry: a
tetrahedron of galaxies is the smallest configuration with a handedness,
and its mirror image cannot be rotated back onto the original. The sign of
$V$ distinguishes the two. The  direction of each spoke is also quantized
into a pixel on the unit sphere (stored as a two-byte
index per edge at graph-construction time, ${\sim}29\%$ of the graph),
enabling efficient grouping by parity at query time without retaining the
original coordinates.  The angular resolution is a runtime
parameter, defaulting to the $8\times32$ grid used throughout this work;
its consequences are quantified below.

\begin{figure}
\centering
\begin{tikzpicture}[scale=0.82,
  gal/.style={circle, fill=black, inner sep=1.5pt},
  spoke/.style={-{Stealth[length=1.8mm]}, semithick},
  far/.style={gray!70, thin}]
  \node[gal, label={[label distance=0pt]below left:$i$}]   (i)  at (0.0, 0.0)  {};
  \node[gal, label={[label distance=0pt]right:$j_1$}]      (j1) at (2.05, 0.35) {};
  \node[gal, label={[label distance=0pt]above:$j_2$}]      (j2) at (0.75, 1.80) {};
  \node[gal, label={[label distance=0pt]below:$j_3$}]      (j3) at (1.25, -0.95) {};
  \draw[far] (j1) -- (j2);
  \draw[far] (j2) -- (j3);
  \draw[far] (j1) -- (j3);
  \draw[spoke] (i) -- (j1) node[midway, below=0.5pt, font=\scriptsize] {$\mathbf{r}_1$};
  \draw[spoke] (i) -- (j2) node[midway, left=0.5pt,  font=\scriptsize] {$\mathbf{r}_2$};
  \draw[spoke] (i) -- (j3) node[midway, left=0.5pt,  font=\scriptsize] {$\mathbf{r}_3$};
  \node[font=\small] at (1.05, -1.85)
    {$V = (\mathbf{r}_1\!\times\!\mathbf{r}_2)\cdot\mathbf{r}_3 > 0$};
  \draw[dash dot, gray] (3.3, -2.15) -- (3.3, 2.25);
  \node[gray, font=\scriptsize, rotate=90] at (3.55, 0.05) {mirror plane};
  \node[gal, label={[label distance=0pt]below right:$i$}]  (mi)  at (6.6, 0.0)  {};
  \node[gal, label={[label distance=0pt]left:$j_1$}]       (mj1) at (4.55, 0.35) {};
  \node[gal, label={[label distance=0pt]above:$j_2$}]      (mj2) at (5.85, 1.80) {};
  \node[gal, label={[label distance=0pt]below:$j_3$}]      (mj3) at (5.35, -0.95) {};
  \draw[far] (mj1) -- (mj2);
  \draw[far] (mj2) -- (mj3);
  \draw[far] (mj1) -- (mj3);
  \draw[spoke] (mi) -- (mj1) node[midway, below=0.5pt, font=\scriptsize] {$\mathbf{r}_1$};
  \draw[spoke] (mi) -- (mj2) node[midway, right=0.5pt, font=\scriptsize] {$\mathbf{r}_2$};
  \draw[spoke] (mi) -- (mj3) node[midway, right=0.5pt, font=\scriptsize] {$\mathbf{r}_3$};
  \node[font=\small] at (5.55, -1.85)
    {$V = (\mathbf{r}_1\!\times\!\mathbf{r}_2)\cdot\mathbf{r}_3 < 0$};
\end{tikzpicture}
\caption{This illustration shows the chirality of a galaxy 4-tuple. A
tetrahedron $(i, j_1, j_2, j_3)$ is the smallest configuration of points
with a handedness, fixed by the sign of the  signed volume $V$
(Equation~\ref{eq:volume}). Its mirror image (right) has identical pair
separations, {\em i.e.} the same six edge bins, but opposite $V$, and no
rotation maps one onto the other. \texttt{GRAMSCI} accumulates the two
handednesses with opposite signs in the parity-odd channel $\zeta^{-}$, which therefore
vanishes for any parity-symmetric field.
\label{fig:chirality}}
\end{figure}

Tetrahedra whose pixelized direction vectors give a mathematically zero
signed volume (repeated pixels, coplanar pixel triples) have no defined
chirality, and assigning them a sign from the
floating-point residue of Equation~(\ref{eq:volume}) would contaminate
$\zeta^{-}$ with compiler- and hardware-dependent noise. \texttt{GRAMSCI}
therefore
assigns degenerate tetrahedra ($|V| < 10^{-9}$) zero weight in the odd
channel; the even channel is unaffected.
Both the discarded fraction and the angular error of the
tetrahedra that are retained fall steeply with the resolution of the
direction grid.  Measured in the estimator's own quadruplet measure,
$6.8\%$ of tetrahedra are pixel-degenerate on a coarse $4\times16$ grid
but only $1.4\%$ on the $8\times32$ grid used here (the default), and
$0.1\%$ at
$23\times91$.  The residual angular error matters only for tetrahedra whose
three unit spoke directions are nearly coplanar, whose recovered odd
amplitude a coarse grid attenuates; at $8\times32$ that attenuation is below
the per-cent level even for the most degenerate shapes we inject
(\S\ref{sec:paritytest}).  The resolution is set at run time, and the code
can alternatively evaluate $V$ directly from the galaxy positions, removing
the discretization entirely at a ${\sim}7\%$ cost in query time relative
to the $8\times32$ grid.

The parity-odd 4pCF $\zeta^{-}$ vanishes for any statistically
parity-symmetric field; a non-zero measurement signals parity violation
\citep{cahn23}, and such measurements directly
constrain parity-violating inflationary scenarios \citep{cabass23}.
Searches for such a signal in BOSS galaxy catalogues
\citep{hou23,philcox22parity}, and most recently in DESI
\citep{slepian25,hou25}, have generated significant
interest, and independent configuration-space
measurements provide a direct cross-check on spherical-harmonic-based
analyses.

\subsection{Connected 4pCF: the disconnected subtraction}
\label{sec:connected}

The raw four-point estimator $\zeta^{(4)} = \mathrm{NNNN}/\mathrm{RRRR}$
measures the {\em total} 4pCF, which contains both the connected
(genuine four-point, trispectrum) signal and a {\em disconnected}
contribution built from products of two-point functions.  For a Gaussian
field the four-point function reduces entirely to this disconnected part
through Wick's theorem, so the connected 4pCF isolates the non-Gaussian
information and is the quantity of interest for primordial-trispectrum and
parity studies \citep{sefusatti07}.  The original code \citep{sabiu19}
reported only the total $\zeta^{(4)}$; GRAMSCI~v2 additionally returns the
connected function.

For four points labelled $1,2,3,4$ the disconnected part is the sum over
the three ways of partitioning them into two disjoint pairs,
\begin{equation}
  \zeta^{(4)}_{\rm disc} = \xi(r_{12})\,\xi(r_{34})
                         + \xi(r_{13})\,\xi(r_{24})
                         + \xi(r_{14})\,\xi(r_{23}),
  \label{eq:disc}
\end{equation}
where $\xi$ is the two-point correlation function and the $r_{ab}$ are the
six pairwise separations of the tetrahedron.  Equation~(\ref{eq:disc}) is
invariant under the $S_4$ relabelling of the four vertices that defines a
configuration, so it depends only on the binned configuration and is
evaluated once per configuration from the binned $\xi$.

The subtraction therefore operates at the level of the \emph{binned}
two-point function: writing $\bar\xi(b)$ for the estimate in radial bin
$b$, the code evaluates
\begin{equation}
  \hat\zeta^{(4)}_{\rm disc} = \bar\xi(b_{12})\,\bar\xi(b_{34})
                             + \bar\xi(b_{13})\,\bar\xi(b_{24})
                             + \bar\xi(b_{14})\,\bar\xi(b_{23}),
  \label{eq:discbin}
\end{equation}
where $b_{ab}$ is the radial bin of edge $(a,b)$ in the canonical
configuration.  Relative to the exact expectation of the binned
estimator, namely the average of Equation~(\ref{eq:disc}) over all
tetrahedra in the configuration's six-dimensional bin cell weighted by
the same quadruplet measure as the $\mathrm{RRRR}$ normalization, this makes two
approximations: (i) it multiplies bin \emph{averages} of $\xi$ instead of
bin-averaging the \emph{product}, and (ii) each $\bar\xi(b)$ is averaged
with the pair-count ($\mathrm{RR}$) radial weighting, whereas within a
tetrahedron cell the edge-length distribution is set by the joint
quadruplet geometry.  Both are second order in the variation of $\xi$
across a bin.  To quantify them we evaluated the exact
bin-cell average of Equation~(\ref{eq:disc}) for a power law
$\xi \propto r^{-1.8}$, typical of galaxies, with logarithmic bins of
constant width $\Delta r/r$.  The fractional residual of the
disconnected term is $4.7$, $1.9$, $0.6$ and $0.14$ per cent in the
median over configurations for $\Delta r/r = 0.4$, $0.2$, $0.1$ and
$0.05$, so $\Delta r/r \lesssim 0.1$ is needed for the disconnected term
to be accurate to about 1 per cent in most configurations.
Because the disconnected term
dominates at these separations, the fractional error on the connected
4pCF is correspondingly larger.  Users who need the connected 4pCF to
per-cent accuracy should therefore use narrow bins or, better, compare
with models or mocks to which the same binned subtraction has been
applied.  This is the case for the DESI comparison of
Figure~\ref{fig:desimock4}: its bins ($\Delta r/r \approx 0.2$--$0.4$)
lie outside the per-cent range, but the \textcolor{black}{data and the
EZmocks are processed with the identical estimator and binning, so the
residual largely cancels between them.  The cancellation is not exact.
The bin-cell measure is common to the two (it is set by the survey
window, which they share), but the residual also depends on the
shape of $\xi$ within the bins, so it differs between data and mocks
by an amount set by the difference in their local slopes: for the power
law above, changing the slope by $0.1$ changes the residual by about
6 per cent of itself, i.e.\ by $0.3$ ($0.1$) per cent of the
disconnected term in the median at $\Delta r/r = 0.4$ ($0.2$), with a
90th percentile of $1.2$ ($0.5$) per cent.  The DESI comparison is
therefore a consistency test of the data against the mock ensemble; an
accurate comparison of the connected 4pCF with a model requires either
$\Delta r/r \lesssim 0.1$ or the model to be passed through the same
binned subtraction.}

The required $\xi$ is computed internally on the {\em same} adjacency
graph and radial bins as the 4pCF, making the subtraction self-consistent
with the four-point measurement (identical bin edges, identical pair set,
no extra catalogue pass beyond a single sweep of the graph).  It uses the
same signed-weight estimator as the standalone 2pCF
(\S\ref{sec:graph}): with data weights normalized to sum $+1$ and
random weights to sum $-1$, the sum of $w_i w_j$ over all pairs in a bin
is the Landy--Szalay numerator $DD - 2DR + RR$, so
$\xi = (\text{signed pair sum})/RR$ is already the correlation function
and vanishes for an unclustered field --- no additive constant enters.
The connected 4pCF is then
\begin{equation}
  \zeta^{(4)}_{\rm conn} = \zeta^{(4)} - \zeta^{(4)}_{\rm disc}.
  \label{eq:conn}
\end{equation}
For the parity decomposition (\S\ref{sec:parity}) the disconnected
term is parity-even and is removed only from the even channel; the
parity-odd 4pCF is intrinsically connected.  At the separations probed
here the disconnected term dominates the total by roughly an order of
magnitude, so isolating the connected residual
(Figure~\ref{fig:desimock4}) is essential for any four-point science.

\subsection{Estimator normalization and survey-window effects}
\label{sec:window}

All GRAMSCI statistics derive from one estimator family, the natural
generalisation of \citet{landy93} to $k$ points \citep{szapudi98}: with
data weights normalized to $\sum_i w_i = +1$ and random weights to $-1$
(Section~\ref{sec:graph}), the signed sum over $k$-tuples in a binned
configuration $B$ is divided by the all-random count of the same
configuration,
\begin{equation}
  \hat\zeta^{(k)}(B) \;=\;
  \frac{\big\langle (D-R)^k \big\rangle_B}{\big\langle R^k \big\rangle_B},
  \label{eq:ssk}
\end{equation}
of which the 2pCF, 3pCF, and 4pCF modes are the $k = 2, 3, 4$ cases.  In
the parity-decomposed case both channels share the \emph{same}
parity-even denominator,
\begin{equation}
  \hat\zeta^{\pm}(B) = \frac{\langle (D-R)^4 \rangle^{\pm}_B}
                            {\langle R^4 \rangle_B},
  \label{eq:parnorm}
\end{equation}
where $\langle\cdot\rangle^{-}$ weights every quadruplet by
$\mathrm{sign}(V)$ (Equation~\ref{eq:volume}) and
$\langle\cdot\rangle^{+}$ is unweighted.  The odd-channel random count
$\mathrm{RRRR}^{-} = \langle R^4\rangle^{-}_B$ is \emph{not} used as a
denominator (for an achiral window it vanishes in expectation) but
is accumulated and written alongside every configuration as a
window-chirality diagnostic, used below.

Because the random catalogue carries the full survey selection
(angular mask, radial selection, and the same systematic and FKP weights
as the data), the numerator of Equation~(\ref{eq:ssk}) removes the mean
windowed density at every vertex, and the denominator normalises each
binned configuration by the window's own $k$-tuple measure.  Edge and
selection effects are therefore removed at leading order configuration
by configuration, exactly as in the two-point \citet{landy93} case, and
no window-deconvolution or configuration-mixing treatment is required:
unlike harmonic-basis $N$-point estimators, whose basis functions extend
across the survey so that the window must be deconvolved from the
measured multipoles, the binned configuration-space estimator is local:
every counted tuple lies entirely within the footprint, and the
window enters only through the per-configuration normalization.  What
Equation~(\ref{eq:ssk}) estimates is correspondingly the \emph{windowed}
statistic: the correlation averaged over the bin cell with the survey's
own tuple measure.

Two residual effects, common to every estimator of this family, are
\emph{not} removed by the randoms: the finite-random bias and variance,
which scale inversely with the random density, and the integral
constraint: the weight normalization measures density contrasts about
the sample mean, suppressing correlations on scales approaching the
window size.  Neither is corrected internally; in this paper both are
controlled in the standard configuration-space way, by processing the
mock ensemble of Section~\ref{sec:desi} through the identical estimator,
so that windowed averaging, finite randoms, and the integral constraint
affect data and reference alike.  The connected 4pCF inherits the same
properties: its disconnected term is built from the windowed internal
$\xi$ of the same catalogue (Section~\ref{sec:connected}), so the
subtraction removes the windowed disconnected part self-consistently,
as demonstrated exactly for a box window in Section~\ref{sec:solvable}.

The parity-odd channel needs one further statement: a survey mask is in
general chirally asymmetric, and for a parity-even galaxy field a chiral
window leaks even-channel signal into the odd channel with expectation
$\mathrm{E}[\hat\zeta^{-}] \simeq
(\mathrm{RRRR}^{-}/\mathrm{RRRR})\,\hat\zeta^{+}$ (mixed data--random
terms cancel in expectation in the odd channel).  
The stored ratio $\mathrm{RRRR}^{-}/\mathrm{RRRR}$ therefore bounds the
leakage configuration by configuration, and where it is non-negligible
the remedy is a single subtraction,
$ \hat\zeta^{-} -
(\mathrm{RRRR}^{-}/\mathrm{RRRR})\,\hat\zeta^{+}$.  For the samples
analysed here no such correction is needed.  Measuring the parity 4pCF
on the DESI DR1 LRG catalogues of Section~\ref{sec:desi}
($45 < r < 65\Mpch$, separately for each cap and redshift bin), we find
median values of $|\mathrm{RRRR}^{-}/\mathrm{RRRR}|$ of
$0.6$--$4\times10^{-3}$.  Two features identify these values as the
finite-random noise floor of the diagnostic itself rather than genuine
mask chirality.  First, they are of the same order as the
$4\times10^{-4}$ obtained with uniform randoms in a periodic box, where
the window is exactly achiral and any non-zero ratio can only be shot
noise.  Second, within the survey the ratio is largest for the sparsest
sub-sample, as expected of shot noise, whereas true chirality is a
property of the mask and would not depend on the sampling density.
Even taken at face value, the measured ratios imply an even-to-odd
leakage below half a per cent of the odd channel's per-configuration
statistical fluctuations in every sub-sample, and we therefore leave
the odd channel uncorrected.

\subsection{Python interface}
\label{sec:python}

A lightweight Python package (requiring only NumPy) wraps the compiled
Fortran binary. Input/output is handled automatically through a temporary
directory; the user supplies NumPy arrays and receives structured result
objects:

\begin{lstlisting}[caption={Minimal Python usage.},label=lst:python]
import gramsci, numpy as np

# pos: (N,3) float64,  weights: (N,) float64
r3 = gramsci.compute_3pcf(
        pos_data, w_data, pos_rand, w_rand,
        rmin=1.0, rmax=30.0, nbins=6)
print(r3.zeta)   # shape (n_configs,)
print(r3.NNN, r3.r1, r3.r2, r3.r3)

r4 = gramsci.compute_4pcf(
        pos_data, w_data, pos_rand, w_rand,
        rmin=1.0, rmax=30.0, nbins=3,
        parity=True)
print(r4.zeta_even)  # parity-even 4PCF
print(r4.zeta_odd)   # parity-odd 4PCF
\end{lstlisting}

The package is installable with \texttt{pip install -e .} from the
\texttt{python/} subdirectory of the repository.

\subsection{GPU implementation}
\label{sec:gpu}

The dominant cost in every \texttt{GRAMSCI} query is the enumeration loop over
hub--neighbor tuples (millions of independent hubs, each performing
$O(m^2)$ to $O(m^3)$ work on small sorted integer lists). This structure maps
naturally onto a GPU, and we have ported the complete query engine (3pCF,
equilateral 3pCF, 4pCF, and parity 4pCF) to OpenACC, compiled with
\texttt{nvfortran}. The GPU binary, \texttt{gramsci\_gpu}, accepts the same
command-line options and produces identical output formats to the CPU code.
Graph construction remains on the CPU (OpenMP) and is typically a small
fraction of the total runtime.

\subsubsection{Data layout}
\label{sec:csr}

The CPU code stores adjacency lists as per-node allocatable arrays, a
structure that cannot be mapped to device memory. The GPU path flattens the
graph into a compressed-sparse-row (CSR) form: a 64-bit offset array
$\mathtt{ptr}(1{:}N{+}1)$ and flat edge arrays holding the neighbor index
(32-bit) and radial-bin index (8-bit), giving 5 bytes per edge, plus a
two-byte direction-pixel index for parity queries. Edge counts routinely exceed
$2^{31}$ for BAO-scale work (\S\ref{sec:performance}), so all edge indexing
is 64-bit. During the flattening, each jagged row is freed as it is copied
(with the freed pages explicitly returned to the operating system), holding
peak host memory to approximately one live copy of the graph.

\subsubsection{Kernel design}
\label{sec:kernels}

The kernels are written against OpenACC's parallelism hierarchy, which the
compiler maps onto the hardware: a \emph{gang} is an independent,    
coarse-grained work unit (CUDA thread block) while the
\emph{vector} lanes within a gang are its fine-grained threads.  On NVIDIA
GPUs these threads are executed in lockstep in groups of 32 called a
\emph{warp}, which is the granularity at which the hardware coalesces
neighbouring memory accesses into one transaction and at which divergent
work between lanes is paid for.  The rules below are, in effect, about
keeping each warp reading one small shared list rather than scattering
across the graph.

Three design rules, found empirically, determine GPU performance:

{\em One thread block per hub, vector lanes on the inner loop.} Each hub is
assigned to one OpenACC gang (CUDA thread block); the innermost neighbor
loop is spread across the gang's vector lanes. For a fixed hub--spoke pair,
all lanes then search or read the {\em same} adjacency list, so a warp's
memory traffic is confined to one small, cache-resident segment. Letting the
compiler choose the mapping instead places a different hub on every vector
lane, producing fully divergent memory access; we measured this na\"ive
mapping to be $6\times$ slower than the 64-thread CPU code, versus
$2.6\times$ faster for the explicit mapping.

{\em Per-hub scratch via gang slots.} The 4pCF kernels precompute, per hub,
a local connectivity matrix $\mathtt{lmat}(k_2,k_1)$ holding the bin index
of the edge between neighbors $k_1<k_2$, reducing the innermost tetrahedron
loop to $O(1)$ array lookups: $C(m,2)$ searches replace the original
$3\,C(m,3)$. This matrix (up to several hundred kilobytes for dense graphs)
cannot be a private array, since OpenACC privatizes arrays per vector lane,
so each gang owns a slice of a global scratch array indexed by an explicit
gang-slot loop. The scratch extent is sized at runtime from the longest
adjacency row of the actual graph.
We refer to this gang-slotted array as the
\emph{connectivity scratch}; its total footprint is the square of the
longest adjacency row times the number of resident gangs, and it
dominates the four-point kernels' device memory at large $\rmax$
(Section~\ref{sec:resource}).

{\em Atomic accumulation into slot-strided partials.} Vector lanes within a
gang race on shared accumulators, and compiler-generated array reductions
across nested gang/vector loops proved unreliable or slow. Counts are
instead accumulated with hardware floating-point atomics into per-slot
partial arrays (one slot per gang, or hash-strided over $\sim$4k slots),
which are summed on the host. Atomic contention is negligible because
concurrently resident gangs own distinct slots.

\subsubsection{Out-of-core operation}
\label{sec:chunking}

At BAO scales a survey-sized graph exceeds any single GPU's memory: the
DESI DR1 LRG North+South sample at $\rmax = 150\Mpch$ produces
$9.0\times10^9$ edges (45\,GB of CSR). The query kernels therefore support
a chunked mode, selected automatically at runtime when the graph exceeds
the free device memory.

The kernel needs two kinds of row access: the hub's own row, and the row of
any neighbor being searched. The edge arrays are split into $W$ row-aligned
windows, and the query runs as a tile loop over (hub-window, search-window)
combinations with only the active windows resident on the device. A triangle
at hub $i$ with spoke pair $(j_1, j_2)$ is processed exactly once, in the
tile whose hub window holds $i$'s row and whose search window holds $j_1$'s
row, and tiles skip foreign spokes with a single integer comparison. For the
4pCF kernels, whose connectivity matrix draws on two different neighbor
rows, tiles run over {\em pairs} of search windows $(c_1 \le c_2)$; the
sorted adjacency invariant guarantees each tetrahedron lands in exactly one
tile. Window sizes are derived at runtime from free device memory and can be
overridden through an environment variable, which also provides a convenient
way to exercise the chunked code paths on small test data.

The measured cost of chunking is modest: the $9.0\times10^9$-edge DESI
measurement above ran in 28 minutes on a 24\,GB RTX 3090~Ti, roughly 20\%
slower than the extrapolated cost of a hypothetical all-resident run, the
difference being PCIe re-transfers of the windows.

The out-of-core scheme thus removes the \emph{memory} barrier to
BAO-scale work for every statistic, the 4pCF included.  For the 4pCF,
however, the binding constraint at the acoustic scale is enumeration
time rather than memory: the $\rmax^{9}$ growth of the tetrahedron count
excludes a full-ball $\rmax = 150\Mpch$ measurement of the complete
$1.5$M-galaxy LRG sample on a single GPU, but within a DESI redshift bin
the complete four-point configuration space is directly measurable
through the acoustic scale --- $\rmax = 130\Mpch$ in a few hours and
$\rmax = 150\Mpch$ in under seven hours on one card
(Section~\ref{sec:desi4pcf} and Table~\ref{tab:resource}).

\section{Performance Benchmarks}
\label{sec:performance}

\subsection{Test catalogue}

All CPU benchmarks used a Gaussian mock with $N_g = 207\,246$ galaxies and
$N_r = 500\,000$ randoms in a periodic box. Timings are wall-clock seconds
on a single node with 4 OpenMP threads, divided by thread count to give
effective single-core seconds for fair comparison. Radial bins are equally
spaced from $r_\mathrm{min} = 1\Mpch$ to $\rmax$.

Benchmarks were run on a shared (but lightly loaded) workstation rather than
a dedicated node. To quantify the resulting timing noise, several
configurations were measured repeatedly at different times during the
campaign: CPU timings reproduce to $0.6$--$1.1\%$ and GPU timings to
${\lesssim}5\%$, far below the effects reported here. In addition, the
headline CPU$\times$64 measurements and the v1 lineage anchor were
re-measured on the same node in a quiescent state after the measurement
campaigns completed, reproducing the stored values to $\le 2.1\%$; the
thread-scaling sweep of Figure~\ref{fig:threads} was re-run in full under
the same quiescent conditions (and is the version shown), reproducing the
parallel efficiencies exactly.

All benchmarks in this paper compare implementations of the same estimator
within \texttt{GRAMSCI}, with identical binning, identical outputs and identical
hardware, so every quoted speedup is directly attributable to the
algorithmic or architectural change being tested. We deliberately make no
cross-code timing claims: the publicly available $N$-point codes
(\S\ref{sec:intro}) compute related but distinct quantities (projected
versus three-dimensional statistics, harmonic-basis multipoles versus binned
configuration-space functions), and timing comparisons across different
output products are not meaningful.

\subsection{Merge-walk speedup}

Table~\ref{tab:mergewalk} summarizes the CPU results. The speedup grows with
$\rmax$ because the mean neighbor count $m \propto \rmax^3$ and the savings
per inner-loop step scale as $\log_2 m$. At $\rmax = 55\Mpch$
($m \approx 700$) the 3pCF achieves $2\times$; at $\rmax = 40\Mpch$ the 4pCF
achieves $3.8\times$ because two independent binary searches were replaced
by a single three-pointer walk.

\begin{deluxetable}{lrrrr}
\tabletypesize{\footnotesize}
\tablewidth{0pt}
\tablecaption{Wall-clock speedup of the merge-walk over binary search.
\label{tab:mergewalk}}
\tablehead{
  \colhead{Kernel} & \colhead{$\rmax$} & \colhead{Bsearch} &
  \colhead{Merge-walk} & \colhead{Speedup} \\
  \colhead{} & \colhead{[$h^{-1}$Mpc]} & \colhead{[s]} &
  \colhead{[s]} & \colhead{}
}
\startdata
3pCF          & 30 &   6.6 &   4.5 & 1.46$\times$ \\
3pCF          & 40 &    34 &    21 & 1.65$\times$ \\
3pCF          & 55 &   223 &   109 & 2.04$\times$ \\
4pCF          & 30 &  61.6 &  18.8 & 3.27$\times$ \\
4pCF          & 40 & 666.7 & 174.1 & 3.83$\times$ \\
4pCF (parity) & 30 &  69.2 &  24.8 & 2.79$\times$ \\
\enddata
\tablecomments{Single-node, 4 OpenMP threads; times divided by thread
  count. The 4pCF gain is larger than the 3pCF gain because the original
  code had two binary searches in the innermost loop.}
\end{deluxetable}

\subsection{Three code generations}
\label{sec:gpubench}

\begin{figure*}[t!]
\centering
\includegraphics[width=\textwidth]{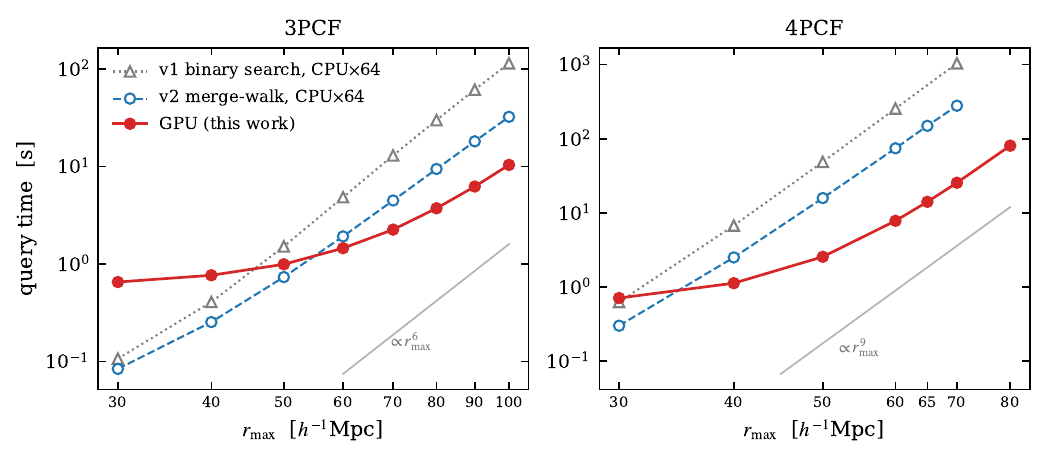}
\caption{This shows the query time versus $\rmax$ for the three generations
of \texttt{GRAMSCI}: the original binary-search kernels (v1, gray triangles), the
merge-walk (v2, open blue circles), and the GPU port (red), for the 3pCF
(left) and 4pCF (right) on the test catalogue ($0.7$M points). Both CPU
generations use all 64 threads. Gray guides show the expected
$r_{\rm max}^{6}$ and $r_{\rm max}^{9}$ scalings of the pair- and
triple-enumeration work. The GPU pays a fixed ${\sim}0.7$\,s transfer cost
and overtakes the CPU at $\rmax \approx 55$ ($35$)$\Mpch$ for the 3pCF
(4pCF).
\label{fig:scaling}}
\end{figure*}

Figure~\ref{fig:scaling} shows the central benchmark of this paper: query
time as a function of $\rmax$ for the three generations of the code (the v1
binary-search kernels, the v2 merge-walk, and the GPU port), all computing
identical binned counts, with both CPU generations using all 64 threads. The merge-walk advantage measured at 4 threads
(Table~\ref{tab:mergewalk}) persists at full thread count ($3.2\times$ for
the 3pCF and $3.7\times$ for the 4pCF at the largest common scales). The GPU
curves carry a fixed ${\sim}0.7$\,s offset (PCIe transfers and kernel
launch), so the CPU wins below a crossover scale ($\rmax \approx 55\Mpch$
for the 3pCF, ${\approx}35\Mpch$ for the 4pCF) beyond which the GPU pulls
away along the same $r_{\rm max}^{6}$ and $r_{\rm max}^{9}$ power laws. At
$\rmax = 70\Mpch$ the 4pCF lineage reads $1044\,\mathrm{s} \to
280\,\mathrm{s} \to 26\,\mathrm{s}$: a cumulative $41\times$ from the
original published kernel to the GPU on the same node.
Table~\ref{tab:gpu} adds measurements on the DESI DR1 LRG sample
(\S\ref{sec:desi}). In all cases the binned counts agree exactly with the
CPU reference.

\subsection{Parallel scaling}
\label{sec:threads}

\begin{figure}
\centering
\includegraphics[width=\columnwidth]{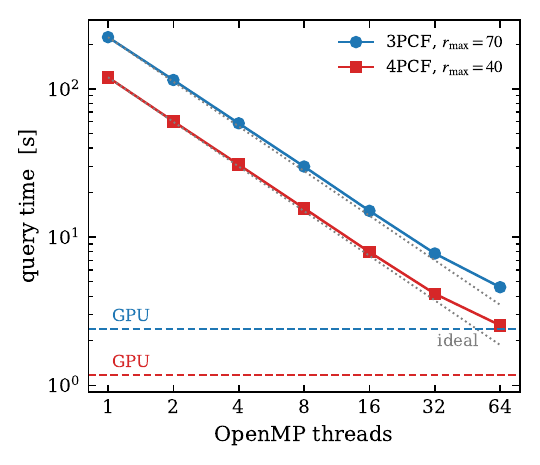}
\caption{This shows the OpenMP strong scaling of the CPU query kernels (3pCF
at $\rmax{=}70\Mpch$, 4pCF at $\rmax{=}40\Mpch$; dotted guides show ideal
scaling). Dashed horizontal lines mark the single-GPU times for the same
problems.
\label{fig:threads}}
\end{figure}

We now investigate the performance of the query kernels with increasing
parallelization. Figure~\ref{fig:threads} shows OpenMP strong scaling of the
v2 CPU kernels from 1 to 64 threads, with the GPU time as a horizontal
reference. Both
kernels scale with $76\%$ (3pCF) and $74\%$ (4pCF) efficiency at 64 threads.
The roll-off reflects the \texttt{reduction}-clause accumulator copies and
dynamic-schedule load imbalance from the long tail of high-multiplicity
hubs. On these problems the single GPU matches roughly $95$--$100$ CPU cores
of equivalent throughput.

\subsection{Density dependence}
\label{sec:density}

\begin{figure}
\centering
\includegraphics[width=\columnwidth]{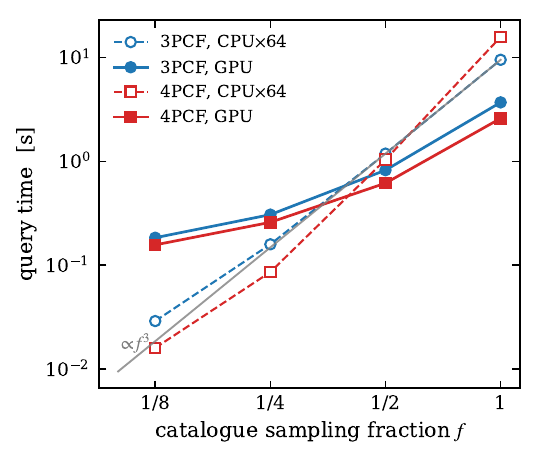}
\caption{This shows the query time versus catalogue sampling fraction $f$ at
fixed $\rmax$ (3pCF at $80\Mpch$, 4pCF at $50\Mpch$). The gray guide shows
$t \propto f^{3}$. The GPU overtakes the 64-thread CPU near $f \approx 0.4$
and its advantage grows with density.
\label{fig:density}}
\end{figure}

Figure~\ref{fig:density} isolates the dependence on sample density by
subsampling the test catalogue (galaxies and randoms together) at fixed
$\rmax$. Query time follows the expected $t \propto f^{3}$ for the 3pCF
(hubs $\propto f$, pairs per hub $\propto f^{2}$) with a steeper $f^{3.3}$
for the 4pCF. The GPU--CPU crossover sits near $f \approx 0.4$: below it the
fixed transfer cost dominates and the CPU is faster; above it the GPU
advantage grows with density, reaching $2.6\times$ (3pCF) and $6.1\times$
(4pCF) at full density and continuing to widen for denser samples. Together
with Figure~\ref{fig:scaling} this defines the regime where the GPU pays
off: roughly, whenever the CPU query exceeds a few seconds.

\subsection{Out-of-core overhead}
\label{sec:chunkbench}

\begin{figure}
\centering
\includegraphics[width=\columnwidth]{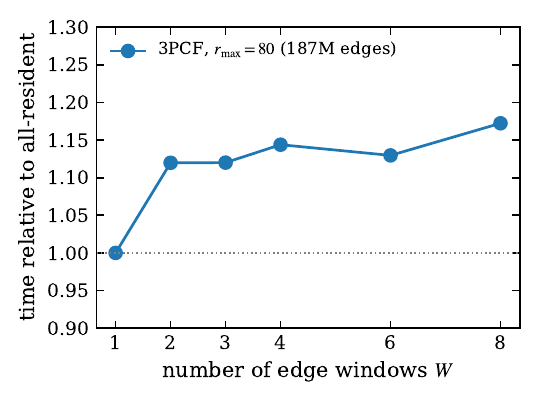}
\caption{This shows the measured out-of-core overhead: query time for a
fixed problem forced through $W$ edge windows, relative to the all-resident
case. The one-time ${\sim}12\%$ cost saturates rather than growing with $W$.
\label{fig:chunks}}
\end{figure}

Figure~\ref{fig:chunks} measures the cost of the window tiling of
\S\ref{sec:chunking} directly, by forcing a fixed problem (3pCF at
$\rmax = 80\Mpch$, $1.9\times10^{8}$ edges) through $W = 1$--$8$ windows.
Going out-of-core costs $12\%$ once, after which the overhead is nearly flat
($17\%$ at $W = 8$): the re-staged windows and repeated hub scans grow only
linearly in $W$ while the dominant pair enumeration is partitioned, not
repeated. This plateau is corroborated at $50\times$ the scale by the DESI
measurement of Table~\ref{tab:gpu}, where the $9.0\times10^{9}$-edge query
at $W = 5$ ran ${\sim}20\%$ above the single-window extrapolation.

\begin{deluxetable}{llrrr}
\tablecaption{GPU query times vs.\ a 64-thread CPU node (RTX 3090~Ti
  vs.\ 64 OpenMP threads).\label{tab:gpu}}
\tablehead{
  \colhead{Kernel} & \colhead{Catalogue} & \colhead{CPU$\times$64} &
  \colhead{GPU} & \colhead{Speedup} \\
  \colhead{} & \colhead{} & \colhead{[s]} & \colhead{[s]} & \colhead{}
}
\startdata
3pCF, $\rmax{=}80$    & test (0.7M)    &   9.4 &   3.6 & 2.6$\times$ \\
4pCF, $\rmax{=}50$    & test (0.7M)    &  15.7 &   2.9 & 5.5$\times$ \\
4pCF, $\rmax{=}65$    & test (0.7M)    & 146.8 &  16.4 & 9.0$\times$ \\
3pCF, $\rmax{=}120$   & DESI LRG (3.0M) & 1030 & 380 & 2.7$\times$ \\
3pCF, $\rmax{=}150$\tablenotemark{a} & DESI LRG (3.0M) & 3718 & 1674 & 2.2$\times$ \\
\enddata
\tablenotetext{a}{Out-of-core: $9.0\times10^9$ edges (45\,GB CSR) on a
  24\,GB device, $5\times5$ window tiles.}
\tablecomments{Query time only; graph construction (CPU, OpenMP) adds
  $\lesssim 10\%$ for the GPU runs. The DESI GPU/CPU speedups are smaller
  than for the test catalogue because the survey graph is dominated by long
  adjacency lists, where the CPU merge-walk is already cache-efficient; the
  GPU advantage there comes chiefly from the out-of-core capability rather
  than raw throughput.}
\end{deluxetable}

\subsection{Expected performance on other hardware}
\label{sec:otherhw}

All GPU timings in this paper were obtained on a single consumer card
(NVIDIA RTX 3090~Ti: Ampere, 24\,GB, ${\sim}1$\,TB\,s$^{-1}$ memory
bandwidth). The query kernels are bound by memory access (sorted-list traversals,
byte-wide bin loads, and table lookups) rather than by floating-point
throughput: the only FP64 work is a few multiplications
and one atomic addition per accepted tuple. Performance should therefore
track {\em memory bandwidth} rather than FP64 capability, and the large FP64
advantage of data-center GPUs is largely irrelevant here. On this basis we
expect roughly $2\times$ on A100-class hardware (2.0\,TB\,s$^{-1}$) and
$3$--$3.5\times$ on H100-class hardware (3.4\,TB\,s$^{-1}$), with the caveat
that occupancy and atomic-throughput differences between architectures are
not captured by this single-axis estimate. Device memory capacity changes
behavior qualitatively rather than quantitatively: an 80\,GB card raises the
single-pass ceiling to ${\sim}1.5\times10^{10}$ edges, so the largest
measurement in this paper would fit without tiling, and eliminates the
out-of-core overhead of \S\ref{sec:chunking}, while NVLink-class host links
shrink the transfer component of that overhead where tiling is still
required. These are expectations, not measurements. Finally, the tile
decomposition of \S\ref{sec:chunking} partitions work by hub window with no
communication between tiles, so a multi-GPU extension, one hub window per
device, is straightforward in principle; we leave its implementation to
future work.

\subsection{Resource summary for a DESI-like analysis}
\label{sec:resource}

In Table~\ref{tab:resource} we report what a realistic survey
analysis costs, for every statistic the code computes, on one DESI DR1
LRG catalogue (the SGC in the redshift bin $0.4 < z \le 0.6$, used in
Section~\ref{sec:desi}).  Each row lists the graph-construction time,
the query time on the 64-thread CPU node and on the RTX~3090~Ti, and
the memory footprint of the graph; every entry is measured except the
flagged 4pCF CPU entries (scaled from directly measured CPU runs at
smaller $\rmax$ on the same catalogue) and the parity-decomposed GPU
time at $\rmax = 150\Mpch$, whose device footprint exceeds the card and
which is scaled from the measured even-channel run at the same scale.  Three structural facts the table makes visible: graph
construction is never the bottleneck (seconds, always ${\lesssim}10\%$
of query time except for the 2pCF, where the build is the entire cost);
the parity decomposition costs $1.78\times$ the even-channel query on
the GPU at identical binning (branch divergence is paid at warp
granularity); and the complete four-point configuration space is now
directly measured through the acoustic scale on a single card: the
$5$--$130\Mpch$ runs fit resident with a $10.1$\,GB connectivity
scratch and no tiling, and the even-channel $5$--$150\Mpch$ run fits
resident at $23$\,GB in $6.7$\,h.  (The connectivity scratch is the per-gang
$\mathtt{lmat}$ buffer of Section~\ref{sec:kernels}, which caches the
edge bin between every pair of the current hub's neighbors; only the
four-point kernels allocate it.)  What limits the reach is enumeration
time together with that scratch, which grows as the square of the
largest adjacency list, rather than the graph itself: at $\rmax = 150\Mpch$
the even-channel run fills the 24\,GB card almost entirely, and the
parity graph, larger by its direction index, pushes the total to
${\sim}25$\,GB, while the graph itself remains a modest few GB.
Unlike the graph, the scratch cannot be tiled away: it is working
memory for the hub currently being processed, and a hub's adjacency
list is the same in every window, so the out-of-core scheme of
Section~\ref{sec:chunking} streams the edges but leaves the scratch
requirement unchanged.

\begin{deluxetable*}{lrrrrrr}

\tablecaption{Resource requirements for a DESI DR1 LRG analysis,
for every statistic the code computes and at the acoustic scale, all on one
catalogue: the South Galactic Cap in the DESI redshift bin
$0.4 < z \le 0.6$ ($160\,150$ galaxies with an equal number of randoms).  Every entry is measured except where flagged.
\label{tab:resource}}
\tablehead{
  \colhead{Statistic} & \colhead{$r$ range} &
  \colhead{Graph} & \colhead{CPU$\times$64} & \colhead{GPU} &
  \colhead{RAM} & \colhead{GPU mem.} \\
  \colhead{} & \colhead{[$h^{-1}$Mpc]} & \colhead{[s]} &
  \colhead{[s]} & \colhead{[s]} & \colhead{[GB]} & \colhead{[GB]}
}
\startdata
2pCF          & 5--150 & 20 & 0.1 & 3.4\tablenotemark{a} & 3.6 & 4 \\
3pCF          & 5--150 & 20 & 231 & 82 & 3.6 & 4 \\
4pCF          & 5--130 & 12 & $6.6\times10^{4}$ & 6507 & 2.5 & 13 \\
4pCF (parity) & 5--130 & 15 & $8.6\times10^{4}$\tablenotemark{b} & 11\,581 & 3.3 & 14 \\
4pCF          & 5--150 & 20 & $2.4\times10^{5}$\tablenotemark{b} & $2.4\times10^{4}$ & 3.6 & 23 \\
4pCF (parity) & 5--150 & 22 & $3.1\times10^{5}$\tablenotemark{b} & $4.2\times10^{4}$\tablenotemark{b} & 4.9 & ${\sim}25$ \\
\enddata
\tablenotetext{a}{The 2pCF query is a negligible fraction of the graph
build on either architecture (the GPU entry is dominated by device
transfer), so graph construction is the entire cost.}
\tablenotetext{b}{Scaled from directly measured runs: the CPU entries from measured CPU runs at smaller $\rmax$ on the same catalogue; the parity GPU entry at $\rmax = 150\Mpch$ from the measured even-channel run at the same scale using the measured $1.78\times$ parity overhead, its ${\sim}25$\,GB footprint exceeding the 24\,GB device.}

\tablecomments{RAM is the host
memory of a CPU run --- essentially the stored half-graph, and the graph is identical for every even
statistic at the same $r$ range.  GPU mem.\ is the total device
footprint: the same graph plus the connectivity scratch of
Section~\ref{sec:kernels}.
}
\end{deluxetable*}

\subsection{Correctness verification}
\label{sec:correctness}

Both implementations are run on the same catalogue and compared bin by bin.
All four GPU kernels (3pCF, equilateral 3pCF, 4pCF, parity 4pCF) were
validated against the CPU reference in both single-pass and chunked modes
(eight combinations in all), with exact agreement of all binned counts. This
agreement also holds at production scale: the DESI LRG 3pCF of
Table~\ref{tab:gpu}, computed on the GPU in chunked mode over
$9.0\times10^9$ edges (680 binned configurations), reproduces the 64-thread
CPU result to a maximum relative difference of $7\times10^{-7}$ in the
counts; this is floating-point rounding from the differing accumulation
order, not algorithmic disagreement. The regression test suite (included in
the repository) checks isolated-pair 2pCF, regular-tetrahedra connected
4pCF, and chiral-tetrahedra parity 4pCF against known analytic values, and
\texttt{tests/benchmark\_3pcf.py} automates the CPU--GPU comparison.

\subsection{Statistical validation on exactly solvable fields}
\label{sec:solvable}

The checks of Section~\ref{sec:correctness} are deterministic: they verify
that the kernels count what they are meant to count.  A referee-grade test
of the \emph{estimators}, in particular of the normalized full 4pCF and
of the internally subtracted connected 4pCF
(Section~\ref{sec:connected}), additionally requires a statistically
defined field whose correlation hierarchy is known exactly, so that the
ensemble mean of the estimator can be compared with an analytic
expectation.  We use two such fields.

\textit{A solvable non-Gaussian process.}  The first is a conservative
(microcanonical, random-permutation) multiplicative cascade
\citep{mandelbrot74} of the $p$-model type \citep{meneveau87}: starting
from a uniform box, each cell is repeatedly split in two per axis and its
mass distributed among the $2^3$ children by a random permutation of a
fixed weight multiset.
Multifractal cascade measures of this
kind have a long history as descriptions of galaxy clustering
\citep[e.g.][]{martinez90,sabiu07}.  Because the full $n$-point correlators of such a cascade
factorize exactly over the merger tree of the dyadic hierarchy, its 2-,
3-, and 4-point functions, including the connected 4pCF, are known
in closed form with no free parameters (all $n$-point statistics of
random cascade processes follow from the transport-operator formalism
of \citealt{greiner98}; Appendix~\ref{app:cascade} records the
expressions used here), while the field itself is strongly
non-Gaussian.  We use $p$-model weights with per-axis mass fractions
$(1\pm m)/2$ and $m = 0.25$, seven cascade levels ($128^3$ leaf cells) in
a $(1\,h^{-1}{\rm Gpc})^3$ box, giving $\xi \simeq 1.0$--$0.6$ over the
measured range: clustering of realistic strength with a connected
four-point term comparable to the disconnected one.  Each realization is
Poisson-sampled to $10^5$ points and measured with the GPU 4pCF and
equilateral 3pCF kernels
against $4\times$ uniform randoms ($r = 10$--$50\Mpch$, 4 radial bins,
224 embeddable configurations of the 276 enumerated); we average 50
realizations.

\begin{figure*}[t!]
\centering
\includegraphics[width=0.82\textwidth]{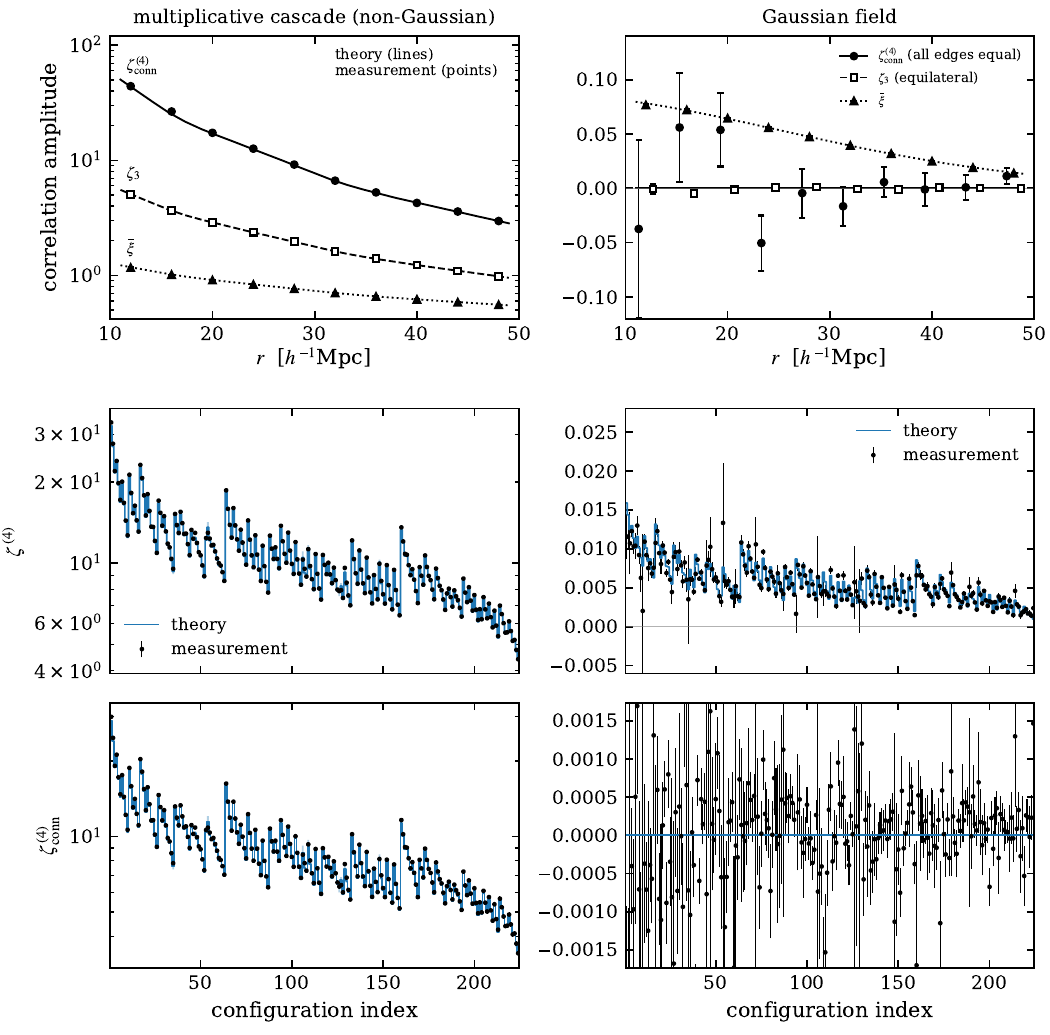}
\caption{Statistical validation of the estimators on fields
with exactly known correlation hierarchies.
\textit{Top panels:} the measured correlation hierarchy on
equal-sided configurations, one panel per field, at finer
($4\Mpch$) radial binning: the bin-averaged 2pCF $\bar\xi(r)$, the
equilateral 3pCF $\zeta_3(r)$, and the connected 4pCF
$\zeta^{(4)}_{\rm conn}(r)$ of the all-edges-equal (regular
tetrahedron) configuration; points are GRAMSCI ensemble means  with
error bars giving the standard error of the mean, curves
the exact closed-form predictions as continuous functions of the side
length (the Gaussian $\zeta_3$ and $\zeta^{(4)}_{\rm conn}$ vanish
identically).
\textit{Lower panels, left column:} conservative multiplicative cascade
(strongly non-Gaussian; 50 Poisson-sampled realizations); \textit{right
column:} Gaussian field (80 realizations), for which the full 4pCF
must equal the Wick pairing sum and the connected 4pCF must vanish.
\textit{First row:} full 4pCF over all 224 configurations, GRAMSCI
ensemble mean (points) against the exact prediction evaluated in the
estimator's own quadruplet measure (band; the error bars are the
standard error of the ensemble mean and the band width is the
Monte-Carlo error of the prediction, which the pulls and
$\chi^2_{\rm diag}$ of Section~\ref{sec:solvable} combine in
quadrature).
\textit{Second row:} connected 4pCF against its exact
counterpart (left) and against zero (right).
\label{fig:mcv}}
\end{figure*}

The analytic expectation is evaluated in the estimator's own measure:
for every binned configuration we Monte-Carlo-sample tetrahedra from its
bin cell inside the box (the same measure that weights $\mathrm{RRRR}$)
and average the exact correlators; the internally
estimated $\bar\xi(b)$ agrees with its exact pair-measure average to
$0.2$ per cent in every bin.  We compare the ensemble mean
$\bar x_i$ of each configuration $i$ with its prediction $x_i^{\rm th}$
through the pull $z_i = (\bar x_i - x_i^{\rm th})/\sigma_i$, where
$\sigma_i^2$ is the squared standard error of the ensemble mean (the
realization-to-realization scatter divided by $N$, the number of
realizations) plus the squared Monte-Carlo error of the prediction, and
quote the diagonal statistic
$\chi^2_{\rm diag} \equiv N_{\rm c}^{-1}\sum_i z_i^2$ over the
$N_{\rm c}$ configurations.  Because every configuration is measured on
the same realizations, the pulls are correlated, and
$\chi^2_{\rm diag}$ does not follow the distribution of a $\chi^2$ with
$N_{\rm c}$ independent degrees of freedom: its null expectation is
$(N-1)/(N-3)$, i.e.\ $1.04$ for $N = 50$ and $1.03$ for $N = 80$, but its
scatter is set by the correlations between configurations.  Across the full configuration space
(Figure~\ref{fig:mcv}, left), the measured full 4pCF matches the exact
prediction with $\chi^2_{\rm diag} = 0.75$, and the
connected 4pCF (the full estimate minus the internal disconnected
subtraction) matches its exact counterpart with
$\chi^2_{\rm diag} = 0.84$; no configuration deviates by more than
$2.0\sigma$ in either statistic.  For the cascade the correlations
are strong: a single mode, a same-sign fluctuation of the overall
amplitude of the field, carries $83\%$ of the realization variance, so
$\chi^2_{\rm diag}$ is effectively a one-mode statistic.  Because the dominant mode is an
amplitude, we also fit a single amplitude $A$ of the prediction to each
realization, with inverse-variance weights, and average over the
ensemble, taking the error from the realization scatter of $A$, which
includes all correlations: $A = 1.012 \pm 0.013$ for the full and
$1.015 \pm 0.016$ for the connected 4pCF, consistent with unity.

\textit{The Gaussian case.}  The second field is the simple Gaussian random field with a specified two-point function
(a $128^3$ grid realization of $P(k) \propto e^{-k^2 r_s^2/2}$ with
$r_s = 25\Mpch$ and cell-scale rms $\sigma_\delta = 0.3$, for which the
truncation of negative densities affects a fraction $4\times10^{-4}$ of
cells and is negligible).  For a Gaussian field Wick's theorem makes the
four-point function purely disconnected, so the full 4pCF must reproduce
the pairing sum of Equation~(\ref{eq:disc}) built from the known $\xi$,
and the connected 4pCF must vanish.  Both hold (Figure~\ref{fig:mcv},
right): the full 4pCF tracks the Wick prediction with
$\chi^2_{\rm diag} = 1.01$ and the connected residual is consistent
with zero ($\chi^2_{\rm diag} = 1.12$) over the same 224 configurations
averaged over 80 realizations, with maximum per-configuration
deviations of $3.4\sigma$ and $3.2\sigma$, as expected among 224
trials.  \textcolor{black}{Both values are close to the null expectation
of $1.03$, and the amplitude fit of the full 4pCF to the Wick
prediction gives $A = 0.997 \pm 0.015$, consistent with unity.}

The top panels of
Figure~\ref{fig:mcv} collect the correlation orders per field on the
equal-sided configurations, remeasured in ten $4\Mpch$ radial bins:
the 2pCF, the equilateral 3pCF, and the all-edges-equal connected
4pCF share one scale axis.  The cascade hierarchy spans
$\bar\xi \simeq 1.2$--$0.6$, $\zeta_3 \simeq 5.0$--$1.0$, and
$\zeta^{(4)}_{\rm conn} \simeq 44$--$3.0$ over the measured range
(the ordering reflects the hierarchical amplitudes
$S_3 \approx 3.5$ and $S_4 \approx 25$ acting on $\xi \gtrsim 0.6$),
and each order rides its closed-form prediction:
$\chi^2_{\rm diag} = 0.82$, $0.53$, and $0.56$ for
$\bar\xi$, $\zeta_3$, and $\zeta^{(4)}_{\rm conn}$ (at this fine
binning the binned expectation and the continuous curves shown differ
by less than a per cent).  The Gaussian
field reproduces its $\bar\xi$ while both higher orders are null
($\chi^2_{\rm diag} = 0.70$ and $1.14$).

\subsection{Validation of the parity estimator}
\label{sec:paritytest}

\begin{figure*}[t!]
\centering
\includegraphics[width=0.98\textwidth]{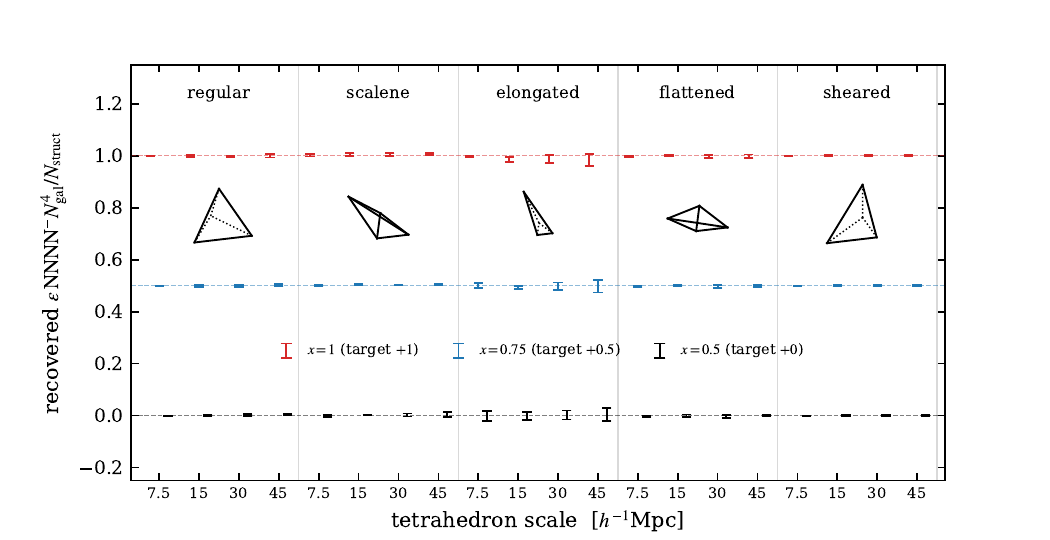}
\caption{Parity-injection recovery across tetrahedron
shapes and scales.  Each column is one (shape, scale) combination
(each family labelled at the top of its block and sketched within it,
hidden edges dotted;
scale in $h^{-1}$Mpc on the axis);
bars span $\pm2\sigma$ of the $\epsilon$-corrected count-level recovered
asymmetry for injected left-handed fractions $x = 1, 0.75, 0.5$ (targets
$+1$, $+0.5$, $0$; dashed lines), over three random realizations with an
independent random orientation per structure.
Measurements use the default $8\times32$ direction grid of
Section~\ref{sec:parity}, at which every family is recovered without
bias at every scale, including the near-degenerate flattened and
sheared ones; a coarser grid attenuates those two.
\label{fig:paritygrid}}
\end{figure*}

The parity-odd channel deserves its own end-to-end test, since it is the
quantity for which subtle sign errors (\S\ref{sec:parity}) are most
consequential. We perform a controlled signal-injection test: catalogues of
$N_{\rm struct} = 500$ isolated scalene tetrahedra of fixed geometry are
generated with a left-handed fraction $x$, embedded in uniform randoms. Each
left-(right-)handed unit contributes $+1$ ($-1$) to the parity-odd count at
the injected edge-bin configuration, and with normalized weights the
count-level prediction is exact: $\mathrm{NNNN}^{-}\,N_{\rm gal}^{4} = N_L -
N_R$, with mixed data--random terms cancelling in the odd channel in
expectation. Across six injection levels spanning $x =
0.5$--$1$ the recovered asymmetry is linear in the injected value and
unbiased at the $0.3\%$ level, with the parity-balanced case ($x=0.5$)
consistent with zero. We note that the same test expressed as the ratio
$\zeta^{-}/\zeta^{+}$ would show a ${\sim}7\%$ offset, not an estimator
bias, but the constant mixed-term contribution to the even channel at this
configuration, which cancels only in the odd channel; count-level
comparisons are the appropriate validation target.

The fixed geometry above establishes linearity in the injected
asymmetry; Figure~\ref{fig:paritygrid} extends the test across
configuration space.  Five shape families are injected at four overall
scales (edge lengths spanning $7.5$--$106\Mpch$) and three left-handed
fractions $x \in \{0.5, 0.75, 1\}$: the regular tetrahedron, the
scalene above, elongated ($\times2.2$) and near-degenerate flattened
($\times0.28$) variants of it, and a sheared tetrahedron.  Every
structure receives an independent random orientation, so the angular
discretization of Section~\ref{sec:parity} is averaged over orientation
as in real data.  The comparison is fully predictive: because the odd
channel's sign convention is defined relative to the canonical vertex
labelling of each binned configuration, the count-level prediction
carries a computable sign $\epsilon = \pm1$, the parity of the $S_4$
permutation relating the injected edge ordering to its canonical form,
which we evaluate by replicating the canonicalization exactly.  With
the default $8\times32$ direction grid used throughout this work the
recovery is unbiased for every family at every scale: at $x = 1$ the
$\epsilon$-corrected recovered asymmetry is $0.999 \pm 0.002$
(regular), $1.004 \pm 0.003$ (scalene), $0.989 \pm 0.009$ (elongated),
$0.999 \pm 0.003$ (flattened) and $1.001 \pm 0.001$ (sheared); the
half-injected case gives $0.4998 \pm 0.0044$ and the balanced case
$0.0008 \pm 0.0051$, with no individual measurement departing from its
target by more than $0.03$.  The near-degenerate families are the
demanding case: their unit-direction spoke volume is small, so a
coarser angular grid attenuates their recovered amplitude
(Section~\ref{sec:parity}).

\begin{figure}
\centering
\includegraphics[width=0.85\columnwidth]{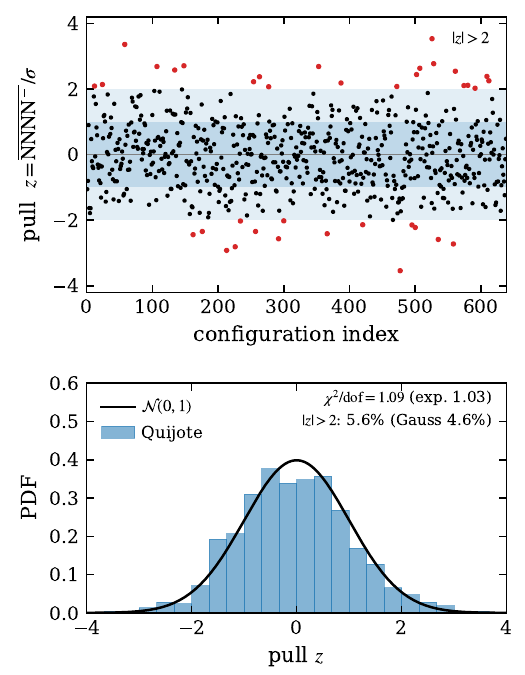}
\caption{Field-level parity null test on the parity-symmetric Quijote
fiducial boxes ($N$-body FoF halos at $z = 0.5$, number density
$\bar n = 1.5\times10^{-4}\,(h/{\rm Mpc})^{3}$).  {\em Top:} the
per-configuration pull $z = \overline{\mathrm{NNNN}^-}/\sigma$ of the
mean parity-odd counts over 66 realizations, with the $\pm1$ and
$\pm2\sigma$ envelopes shaded and the points beyond $2\sigma$
highlighted.  {\em Bottom:} the pull distribution against the unit
Gaussian.  Gravity is parity symmetric, so a correct estimator must
return a null; the measured $\chi^2_{\rm diag} = 1.10$ over 638
configurations (Student-$t$ expectation $1.03$) confirms it.  The same
test without the degenerate-volume rule of \S\ref{sec:parity} is
contaminated by floating-point-signed degenerate tetrahedra.
\label{fig:qparity}}
\end{figure}

The complementary null test exploits the fact that gravity produces no
chirality: on a parity-symmetric cosmological density field the
parity-odd 4pCF must be consistent with zero configuration by
configuration.  We measure it on the Friends-of-Friends halo catalogues
of the Quijote simulation suite \citep{villaescusa20} --- $512^3$
particles in $(1\,h^{-1}{\rm Gpc})^3$ boxes at $z = 0.5$ in the fiducial
cosmology, cut to $\bar n = 1.5\times10^{-4}\,(h/{\rm Mpc})^{3}$ (the
$150\,000$ most massive halos per box) with uniform randoms at 1:1.
Across 66 realizations and 638 well-measured configurations
($\rmax = 65\Mpch$, 5 radial bins) the per-configuration pull
$z = \overline{\mathrm{NNNN}^-}/\sigma$ of the mean parity-odd counts
against zero is distributed as expected for a null
(Figure~\ref{fig:qparity}): $\chi^2_{\rm diag} = 1.10$, with $5.6\%$ of
configurations beyond $2\sigma$ (Gaussian expectation $4.6\%$).  The
appropriate reference value is not unity:
with $N_{\rm m}$ realizations the per-configuration error estimates make
the z-scores Student-$t$ distributed, giving an expected
$\chi^2_{\rm diag} = (N_{\rm m}-1)/(N_{\rm m}-3) = 1.03$\textcolor{black}{;
the off-diagonal covariance between configurations, not included here,
would widen its scatter}.  The same statistic computed without
the degenerate-volume rule of \S\ref{sec:parity} would be
contaminated by floating-point-noise-signed tetrahedra; the passing null
validates that treatment at the field level.

\begin{figure*}[t!]
\centering
\includegraphics[width=0.94\textwidth]{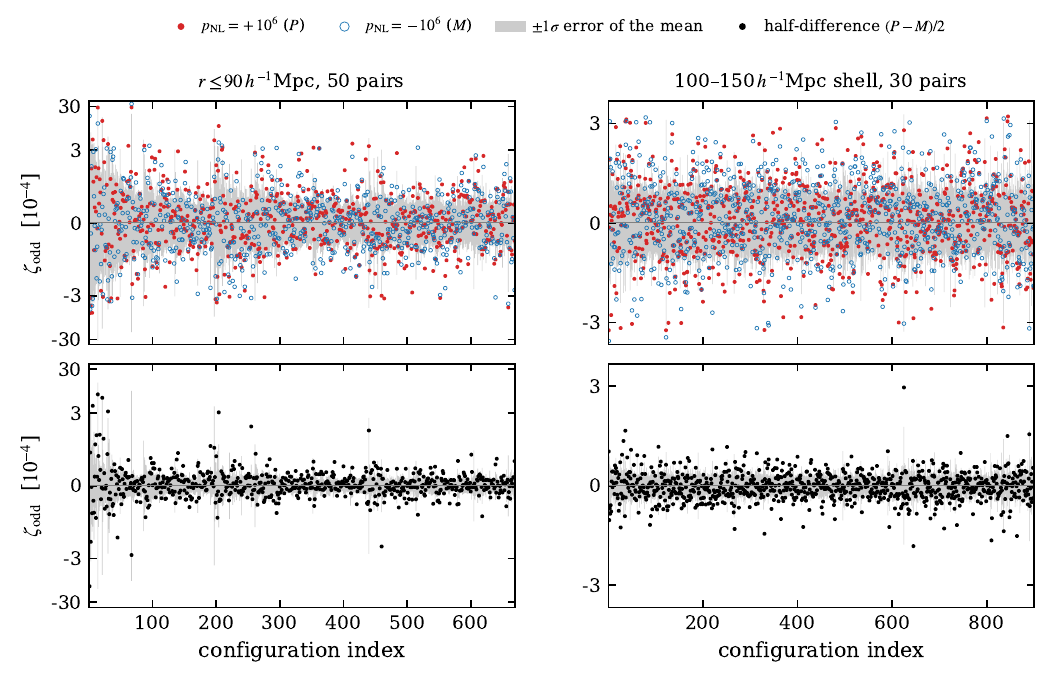}
\caption{The parity-odd 4pCF measured on the Quijote-ODD
suite: simulations with a parity-violating primordial trispectrum
($p_{\rm NL} = \pm10^{6}$) injected into phase-matched, mirrored
initial-condition pairs \citep{coulton23}, for the full configuration
space to $r \le 90\Mpch$ (left; 50 pairs, 669 configurations) and the
$100$--$150\Mpch$ shell (right; 30 pairs, 900 configurations).
\textit{Top:} ensemble-mean $\zeta_{\rm odd}$ for $p_{\rm NL} = +10^6$
($P$, filled) and $-10^6$ ($M$, open) against configuration index, with
the $\pm1\sigma$ error of the mean (band).  \textit{Bottom:} the
half-difference $(P-M)/2$, which cancels the gravitational and
shot-noise fluctuations common to each phase-matched pair and isolates
the injected signal, with its $\pm1\sigma$ error of the mean.  The
vertical axis is linear within $\pm3\times10^{-4}$ and logarithmic
beyond.
\label{fig:oddsuite}}
\end{figure*}

The Quijote suite also provides the opposite regime: simulations in
which parity violation \emph{is} present.  The Quijote-ODD runs
\citep{coulton23} inject a parity-violating primordial trispectrum of
amplitude $p_{\rm NL} = \pm10^{6}$ into otherwise identical,
phase-matched initial conditions, in mirrored pairs.  We measure the
parity 4pCF on the corresponding $z = 1$ FoF halo catalogues (the same
number-density cut and uniform 1:1 randoms as above): 50 matched pairs
over the full configuration space to $r \le 90\Mpch$ (6 radial bins;
669 realisable configurations), and 30 matched pairs with all six edges
in a $100$--$150\Mpch$ shell (900 configurations), the regime where
the injected template, which weights configurations by their signed
volume and peaks on large scales, is strongest, and which the shell
pruning of Section~\ref{sec:desi4pcf} makes affordable.  Because the
pairs are phase-matched (per-pair configuration correlation $+0.81$),
the antisymmetric combination $(P - M)/2$ cancels gravity and shot
noise realization by realization and isolates the injected signal.
Figure~\ref{fig:oddsuite} shows the measured parity-odd
4pCF itself, configuration by configuration, for both signs of
$p_{\rm NL}$ and for their half-difference.  The two signs scatter about
zero within their statistical errors, and the half-difference, whose
error is about three times smaller than that of either sign, is
consistent with zero at both scales: $\chi^2_{\rm diag} = 1.04$ over the full configuration space
(Student-$t$ expectation $1.04$) and $1.02$ in the shell (expectation
$1.07$); the volume-weighted matched filter built from the
injection template itself gives $\langle z^2 \rangle = 1.01$, with the
fattest-volume quartile, where a detection must peak, the lowest of the
four ($0.88$).  The injected large-scale signal \textcolor{black}{is
therefore not detected above} the halo sampling noise at the
configuration-space scales accessible to direct enumeration, even under
\textcolor{black}{the volume-weighted} filter; detecting it is the
province of large-scale harmonic estimators.  \citet{coulton23}
detect the injected parity-odd trispectrum in the $z = 0$ dark-matter
field, using all 500 simulation pairs of the suite and wavenumbers
$k = 0.02$--$0.12\,h\,{\rm Mpc}^{-1}$ that reach the box scale, whereas
our measurement uses 50 (30) pairs, a sparse halo sample, and
separations below $150\Mpch$.  \textcolor{black}{We have not computed the
amplitude that the injected template predicts in these bins, which
depends on its projection onto the measured tetrahedron shapes and
scales and may be strongly suppressed relative to its large-scale
Fourier-space signal.  The null is therefore not a measurement of the
estimator's sensitivity to this template, and should not be read as
one: the injection test of Figure~\ref{fig:paritygrid} demonstrates the
recovery of a localized handedness asymmetry, not the response to a
smooth, large-scale parity-odd trispectrum, whose validation at these
scales would require the template's configuration-space prediction and
is left to future work.}  For the estimator itself
the suite constitutes a second, more stringent field-level null
(any artifact correlated with the initial conditions would survive the
antisymmetric combination), while the positive control, that
\textcolor{black}{localized} configuration-space parity signals are
recovered without bias, is Figure~\ref{fig:paritygrid}.

\section{Application to DESI DR1 LRG}
\label{sec:desi}

\begin{figure*}
\centering
\includegraphics[width=\textwidth]{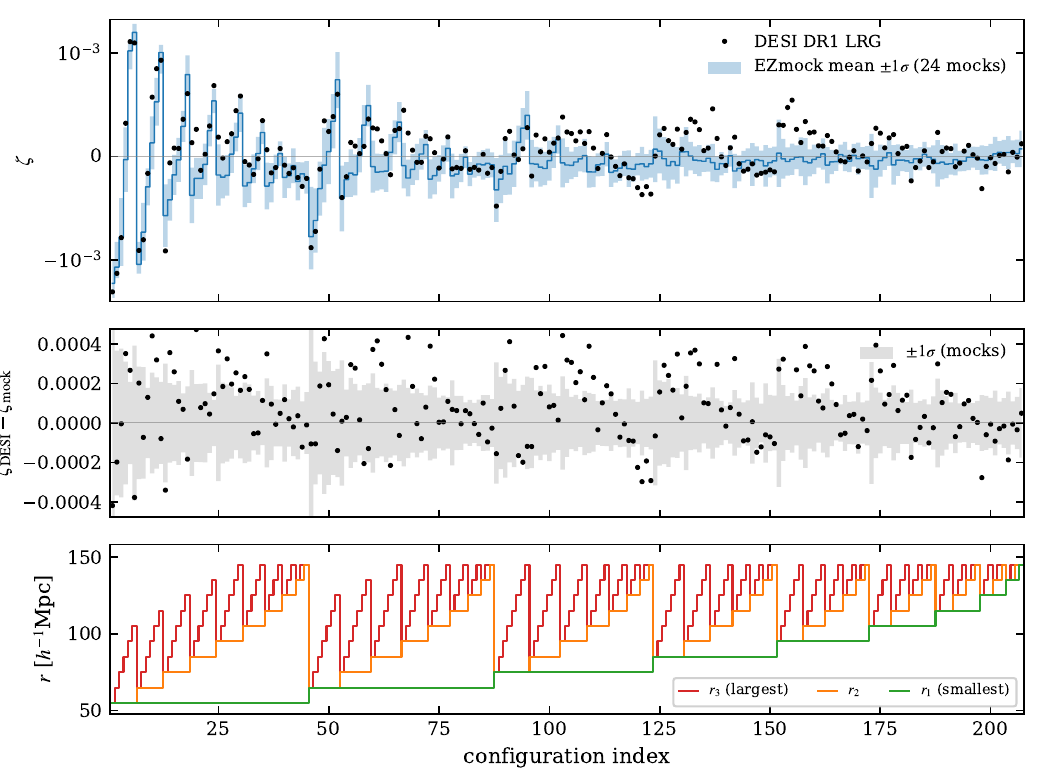}
\caption{DESI DR1 LRG 3pCF versus the EZmock ensemble over all $220$
triangle configurations ($50 < r < 150\Mpch$, 10$\Mpch$ bins; NGC and SGC 
combined at the count level in data and mocks).
{\em Top:} $\zeta$ (symlog axis), DESI points over the mock mean
$\pm 1\sigma$ band (24 realizations).  {\em Middle:} the residual
$\zeta_{\rm DESI} - \bar\zeta_{\rm mock}$ with the $\pm 1\sigma$ mock
band.  {\em Bottom:} a key to the configuration axis --- the three
sorted triangle side-length bins $r_1 \le r_2 \le r_3$ of each
configuration, so a vertical read-off gives the triangle shape.  The
nested staircase reflects the lexicographic ordering of the canonical
configurations.
\label{fig:desimock3}}
\end{figure*}

As a realistic end-to-end application we measure the 2pCF, 3pCF, and
connected 4pCF of the DESI DR1 LRG sample \citep{desi24} ($1.48$M galaxies
NGC, $0.66$M SGC; weights $w \times w_{\rm FKP}$; DESI fiducial cosmology)
with randoms subsampled 1:1, and compare against the EZmock ensemble
\citep{chuang15} processed through the identical pipeline. Galactic caps are
measured independently and combined at the level of raw counts (weighting
each cap's normalized counts by the cube, or square for pair counts, of its
summed data weights). The data cover the redshift range $0.4 < z \le 1.1$.

The combined $r^2\xi(r)$ shows the acoustic peak at $r \simeq 97.5\Mpch$
with the expected dip--peak structure, recovered independently in both caps.
The full-sample 3pCF, measured to $\rmax = 150\Mpch$ in 5$\Mpch$ bins (the
$9.0\times10^{9}$-edge out-of-core measurement of Table~\ref{tab:gpu}),
shows the corresponding BAO feature: in the isosceles slice the acoustic
`bump' appears at $r_3 \simeq 107$--$118\Mpch$, an order of magnitude above
the surrounding baseline, with both Galactic caps showing the rise
independently.

\subsection{3pCF versus the mock ensemble}
\label{sec:desi3pcf}

Figure~\ref{fig:desimock3} compares the full DESI 3pCF against the EZmock
ensemble across all $220$ binned triangle configurations
($50 < r < 150\Mpch$, 10$\Mpch$ bins), with NGC and SGC combined at the 
count level (effective-volume weighting, rebinned exactly from the native 5$\Mpch$ 
measurement).  The top panel
shows $\zeta$ on a symmetric-log axis, which spans the hierarchical
decay from the small-scale configurations (left) to the near-zero
large-scale configurations (right); the middle panel shows the residual
$\zeta_{\rm DESI} - \bar\zeta_{\rm mock}$ against the mock scatter; and
the bottom panel keys the configuration axis to physical scale.  Because
a triangle is specified by three side lengths, this key is three lines:
the sorted side-length bins $r_1 \le r_2 \le r_3$ of each configuration,
whose nested staircase reflects the lexicographic ordering of the
configurations.  Across the $213$ well-measured configurations the data
track the ensemble with $\chi^2_{\rm diag} = 1.4$ (the
off-diagonal configuration covariance, not included here, would change
this), confirming that the configuration-space three-point signal of the
LRG sample is consistent with the $\Lambda$CDM-calibrated mocks across the
full triangle space.

\subsection{4pCF versus the mock ensemble}
\label{sec:desi4pcf}

\begin{figure*}
\centering
\includegraphics[width=\textwidth]{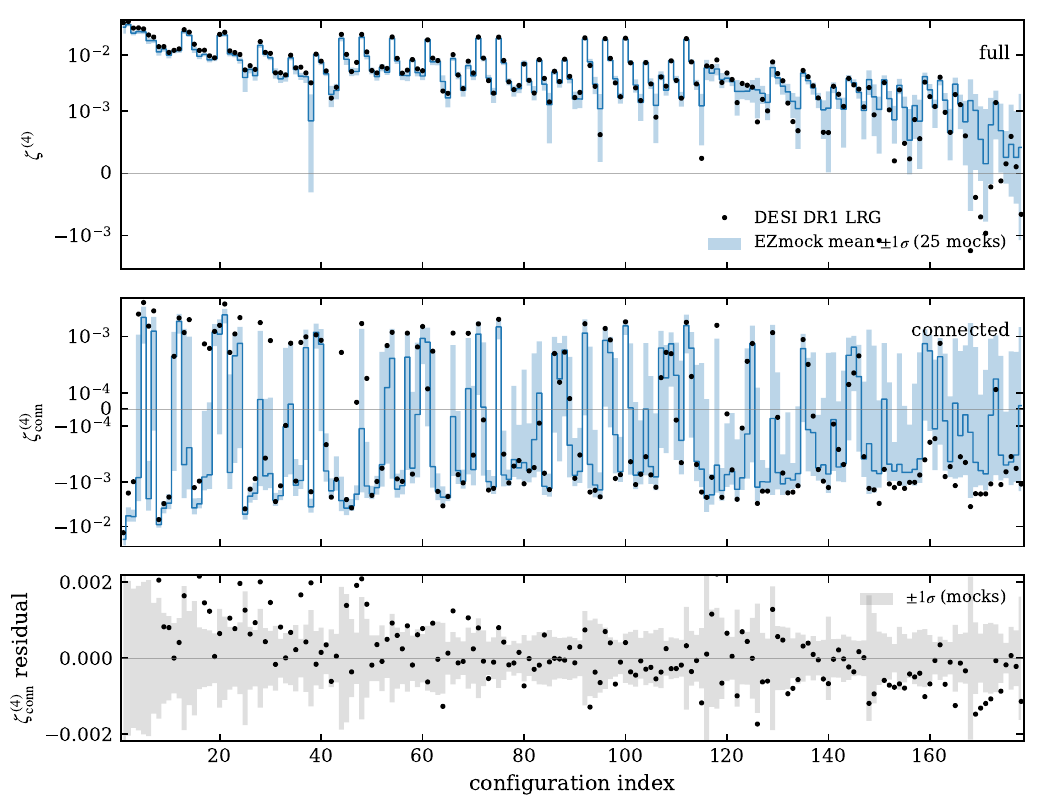}
\caption{DESI DR1 LRG 4pCF versus the EZmock ensemble over the 178
tetrahedral configurations ($20 < r < 65\Mpch$, 4 radial bins; NGC and SGC 
combined at the count level in data and mocks;
configurations that do not form a geometric tetrahedron --- by the
Cayley--Menger criterion on the bin centers --- are excluded).
{\em Top:} the full 4pCF $\zeta^{(4)} = \mathrm{NNNN}/\mathrm{RRRR}$
(symmetric-log axis), dominated by the disconnected Gaussian part.
{\em Middle:} the connected 4pCF
$\zeta^{(4)}_{\rm conn} = \zeta^{(4)} - \zeta^{(4)}_{\rm disc}$ (symlog),
an order of magnitude smaller and sign-changing.  {\em Bottom:} the
connected residual
$\zeta^{(4)}_{\rm conn}(\mathrm{DESI}) - \langle\zeta^{(4)}_{\rm conn}\rangle_{\rm mock}$
with the $\pm 1\sigma$ mock band.  Black points are DESI; the blue band
is the mock ensemble mean $\pm 1\sigma$ (25 realizations).
\label{fig:desimock4}}
\end{figure*}

Figure~\ref{fig:desimock4} extends the comparison to the 4pCF, combining
NGC, SGC  into a single all-LRG measurement at the
count level ($S^4$ effective-volume weighting for the four-point counts,
$S^2$ for the disconnected two-point term).  We show the configurations
with $20 < r < 65\Mpch$ in 4 radial bins whose six edge-length bins form
a realizable tetrahedron: the canonical enumeration also produces
6-tuples that satisfy every face triangle inequality yet are not
embeddable in three dimensions (for example five edges of $26\Mpch$ and
one of $48\Mpch$), which we exclude by requiring a positive
Cayley--Menger determinant on the bin centers, leaving 178 of 276
configurations.

The top panel shows the full 4pCF
$\zeta^{(4)} = \mathrm{NNNN}/\mathrm{RRRR}$, which at these scales is
dominated by the disconnected (Gaussian) contribution --- the sum of
products of pairwise 2pCFs --- and follows a smooth hierarchical decay.
The middle panel shows the connected 4pCF
$\zeta^{(4)}_{\rm conn} = \zeta^{(4)} - \zeta^{(4)}_{\rm disc}$, the
genuine four-point (trispectrum) signal left after subtracting the
disconnected term, which GRAMSCI builds from an internally computed 2pCF;
it is roughly an order of magnitude smaller and changes sign across the
configuration space.  The bottom panel shows the connected residual
against the mock scatter, in which the binning offset of
the disconnected subtraction (Section~\ref{sec:connected}), \textcolor{black}{common
to data and mocks up to the small difference in the shapes of their
2pCFs, cancels; the comparison therefore tests the consistency of the
data with the mock ensemble, not the connected 4pCF against a model}.
The data track the ensemble: the connected
4pCF has $\chi^2_{\rm diag} = 1.15$ (178 configurations; with 25
mocks the per-configuration scatter is itself uncertain at the
${\sim}15\%$ level, and the off-diagonal covariance, not included here,
would change this), consistent with the $\Lambda$CDM-calibrated mocks.

A practical note on cost: the same measurement at $\rmax = 80\Mpch$
takes $2.6$\,h for  the NGC, which is an order of
magnitude beyond the uniform-catalogue scaling of
\S\ref{sec:gpubench}, because the $C(m,3)$ tetrahedron enumeration is 
dominated by the high-multiplicity tail of the neighbor
distribution, which is far heavier in clustered, FKP-weighted survey
data than in quasi-uniform test catalogues ($\langle m^3 \rangle \gg
\langle m \rangle^3$).  The approximate EZmocks, which underpredict
the most extreme small-scale clustering, run ${\sim}4\times$ faster
than the data at identical sample size.  Benchmarks calibrated on
uniform catalogues should therefore be treated as lower bounds for
survey-data 4pCF work.

The same scaling sets how far the four-point measurement reaches.  The
comparison above uses $\rmax = 65\Mpch$ on the combined caps, but the
cost is governed by $N\langle m\rangle^{3}$, so a smaller and sparser
sub-sample buys a great deal of radial range.  Measured on the SGC in the
DESI bin $0.4 < z \le 0.6$ with an equal number of randoms, the complete
configuration space out to $\rmax = 130\Mpch$ takes $1.8$\,h for the even
channel and $3.2$\,h for the parity decomposition on one card, and the
even-channel measurement over the complete configuration space to
$\rmax = 150\Mpch$ takes $6.7$\,h, resident on the same 24\,GB card
(Table~\ref{tab:resource}): the full four-point configuration space is
measured directly through the acoustic scale.  The parity decomposition
at $150\Mpch$ exceeds the card's memory by ${\sim}1$\,GB and is the one
four-point measurement that still requires an 80\,GB device or one of
two further routes.  First, the tile decomposition parallelises over
hub windows with no inter-tile communication, so the wall time falls
linearly with the number of GPUs (Section~\ref{sec:otherhw}).  Second,
because raising
$r_{\rm min}$ prunes the adjacency lists themselves, \emph{shell}
configurations that isolate the large-scale regime are cheaper still: a
parity 4pCF restricted to all six edges in $100$--$150\Mpch$ runs in
${\sim}30$ minutes per $150\,000$-halo $(1\,h^{-1}{\rm Gpc})^3$
catalogue on the same GPU.  Note that raising $r_{\rm min}$ helps only
once it is a substantial fraction of $r_{\max}$: the work scales as
$(r_{\max}^{3} - r_{\rm min}^{3})^{3}$, so trimming $r_{\max}$ is the more
effective lever.


\section{Scientific Applications}
\label{sec:applications}

The new performance regime opens up several science applications that were
previously impractical. We highlight three.

\paragraph{Primordial non-Gaussianity}
The connected 4pCF $\zeta_{\rm conn}$ is a direct configuration-space probe
of the primordial trispectrum. The $9\times$ GPU
speedup makes it feasible to evaluate $\zeta_{\rm conn}$ across the
complete configuration space through the acoustic scale
($\rmax = 150\Mpch$ per DESI redshift bin in under seven GPU-hours;
Table~\ref{tab:resource}), where covariance estimation
requires $\mathcal{O}(10^3)$ mock realizations; the parity-decomposed
$\rmax = 150\Mpch$ ball is reached with the multi-GPU tile decomposition
or the large-scale shell configurations discussed in
Section~\ref{sec:desi4pcf}.

\paragraph{Parity violation}
\citet{hou23} reported a non-zero parity-odd 4pCF in BOSS DR12 CMASS at
$7.1\sigma$ using a spherical-harmonic estimator, with an independent
analysis finding lower significance \citep{philcox22parity};
the first DESI measurements have recently appeared
\citep{slepian25,hou25}, the latter presenting the DESI DR1 LRG
parity-odd 4pCF analysis and finding a signal consistent with zero.
Beyond detection significance, such measurements bound concrete
early-Universe models: \citet{cabass23} used the BOSS parity-odd 4pCF
to constrain a broad family of parity-violating inflationary
scenarios. The \texttt{GRAMSCI}
configuration-space parity estimator provides a complementary cross-check
that is sensitive to the full angular structure of the signal without
assuming a harmonic decomposition. The deterministic treatment of degenerate
tetrahedra (\S\ref{sec:parity}) is  essential here: at the few-percent level
the odd channel is otherwise contaminated by floating-point noise whose sign
depends on the compiler and hardware.

\paragraph{BAO with higher-order statistics}
\texttt{GRAMSCI} supports anisotropic 3pCF and 4pCF (via $\mu$ binning for
redshift-space distortions), enabling joint 2pCF+3pCF+4pCF BAO fits that
extract additional signal-to-noise from the same catalogue.

\section{Conclusions}
\label{sec:summary}

We have presented five additions to \texttt{GRAMSCI}.

We replace the binary-search inner kernel with a two-pointer sorted-set
intersection, the merge-walk, reducing the 3pCF inner loop from $O(m\log m)$
to $O(m)$ and the 4pCF inner loop, which required two binary searches,
correspondingly further. Measured speedups are 1.5--2$\times$ for the 3pCF
and 2.8--3.8$\times$ for the 4pCF at survey-relevant scales. The key
prerequisite, adjacency lists sorted by node index, was already guaranteed
by the kd-tree construction and required no data-structure change.

We present a parity-decomposed 4pCF. A two-byte direction-pixel index stored per edge
at graph-construction time allows the 4pCF to be decomposed into parity-even
and parity-odd channels at modest additional cost
($1.78\times$ the even-channel query on the GPU at survey binning;
Table~\ref{tab:resource}), enabling direct tests
of parity violation. Degenerate (zero-volume) tetrahedra are excluded from
the odd channel deterministically.

We add an internally computed two-point function, evaluated on the same graph
and binning,  which supplies the disconnected (Gaussian) term $\zeta^{(4)}_{\rm disc}$
that can be subtracted from the total 4pCF to return the connected 4pCF at the
cost of a single extra sweep of the graph.  Both
estimators are validated statistically on exactly solvable fields: a
strongly non-Gaussian multiplicative cascade, whose closed-form full and
connected 4pCF they reproduce without bias, and a Gaussian field, where
the full 4pCF matches the Wick pairing sum and the connected 4pCF is
consistent with zero (Section~\ref{sec:solvable}).

We provide a Python interface. A thin NumPy-based wrapper calls the Fortran
engine via subprocess, handling all file I/O automatically and returning
structured result objects.

Finally, and principally, we present a GPU port with out-of-core support.
The complete query engine runs on a single GPU via OpenACC with measured
speedups of $2.6\times$ (3pCF) to $9\times$ (4pCF) over a 64-thread node,
and an automatic window-tiling scheme processes graphs far larger than
device memory ($9\times10^9$ edges demonstrated on a 24\,GB card). All
kernels reproduce the CPU reference exactly.

The code is publicly available at \url{https://github.com/csabiu/Gramsci}.


\section*{Acknowledgments}
We thank the anonymous referee whose comments and suggestions substantially improved this manuscript.

C.G.S. acknowledges support from the Basic Science Research Program 
(2018R1A6A1A06024977) through Korea's NRF funded by the Ministry of Education.

This work was also supported by the UBAI computing resources at the University of Seoul.

This research used data obtained with the Dark Energy Spectroscopic Instrument (DESI). DESI construction and operations is managed by the Lawrence Berkeley National Laboratory. This material is based upon work supported by the U.S. Department of Energy, Office of Science, Office of High-Energy Physics, under Contract No. DE-AC02-05CH11231, and by the National Energy Research Scientific Computing Center, a DOE Office of Science User Facility under the same contract. Additional support for DESI was provided by the U.S. National Science Foundation (NSF), Division of Astronomical Sciences under Contract No. AST-0950945 to the NSF's National Optical-Infrared Astronomy Research Laboratory; the Science and Technology Facilities Council of the United Kingdom; the Gordon and Betty Moore Foundation; the Heising-Simons Foundation; the French Alternative Energies and Atomic Energy Commission (CEA); the National Council of Humanities, Science and Technology of Mexico (CONAHCYT); the Ministry of Science and Innovation of Spain (MICINN), and by the DESI Member Institutions: www.desi.lbl.gov/collaborating-institutions. The DESI collaboration is honored to be permitted to conduct scientific research on I'oligam Du'ag (Kitt Peak), a mountain with particular significance to the Tohono O'odham Nation. Any opinions, findings, and conclusions or recommendations expressed in this material are those of the author(s) and do not necessarily reflect the views of the U.S. National Science Foundation, the U.S. Department of Energy, or any of the listed funding agencies.

This work also uses data from the Quijote simulations \citep{villaescusa20}; we thank the Quijote team for 
making the halo catalogues publicly available. The DESI DR1 EZmock catalogues used here are approximate mock catalogues built on the effective Zel'dovich approximation method \citep{chuang15}, produced for the DESI DR1 large-scale-structure analysis.

Portions of the GRAMSCI v2 code, including
the GPU port and the merge-walk implementation, were developed with the assistance of
Anthropic's Claude large language model, specifically Opus 4.8 and Fable.  All algorithm additions and amendments were reviewed and validated by the author, including against independent
analytic predictions (Sections~\ref{sec:solvable} and
\ref{sec:paritytest}), and the author takes full responsibility for the
code and the scientific content.

\software{Astropy \citep{astropy13, astropy18, astropy22},
          GRAMSCI v2.5 \citep[][code:
          \citealt{sabiu26code}]{sabiu19},
          \texttt{kdtree2} \citep{kennel04},
          Matplotlib \citep{hunter07},
          NumPy \citep{harris20},
          SciPy \citep{virtanen20}}

\section*{Data Availability}

The data underlying this article are from the first data release (DR1) of
the Dark Energy Spectroscopic Instrument, specifically the DR1 LRG
large-scale-structure catalogues and their associated randoms, publicly
available at \url{https://data.desi.lbl.gov/public/dr1}  and described in the DR1
overview paper \citep{desidr1}. The Quijote halo catalogues
\citep{villaescusa20} are public at \url{https://quijote-simulations.readthedocs.io}. 
\texttt{GRAMSCI} is available at
\url{https://github.com/csabiu/Gramsci}; the version
described in this paper corresponds to the tagged release
\texttt{v2.5}, a frozen archive of which is deposited on Zenodo
\citep[DOI 10.5281/zenodo.22091542;][]{sabiu26code}.


\clearpage
\appendix

\section{Closed-form \texorpdfstring{$n$}{n}-point functions of the conservative cascade}
\label{app:cascade}

This appendix records the exact correlators of the conservative
multiplicative cascade used in Section~\ref{sec:solvable}.
The cascade lives on a box subdivided $\Lambda$ times into $2^d$
children per cell ($d = 3$, $\Lambda = 7$ here).  At every subdivision
the parent's mass is distributed among its children by a uniformly
random permutation of the fixed weight multiset $\{w_1, \ldots,
w_{2^d}\}$, $\sum_i w_i = 1$, so mass is conserved cell by cell.  In
terms of the density multipliers $m_i = 2^d w_i$ (unit mean), the leaf
density at a point is the product of the multipliers along its tree
path.

Because the permutations at different parents and different levels are
independent, every moment reduces to \emph{sibling moments}: for $j$
distinct children of one parent carrying $s_1, \ldots, s_j$ powers of
the density,
\begin{equation}
  M_{s_1 \cdots s_j} \equiv
  \big\langle m_{a_1}^{s_1} \cdots m_{a_j}^{s_j} \big\rangle
  = 2^{\,d \sum_i s_i}\,
    \frac{(2^d - j)!}{2^d!}
    \sum_{\substack{i_1, \ldots, i_j \\ \mathrm{distinct}}}
    w_{i_1}^{s_1} \cdots w_{i_j}^{s_j},
  \label{eq:sibmom}
\end{equation}
the sum running over ordered distinct assignments.  Two special cases
recur: $M_q = 2^{d(q-1)} S_q$ with $S_q = \sum_i w_i^q$, and the pair
split $c_2 \equiv M_{11} = 2^{2d} (1 - S_2) / [2^d (2^d - 1)]$, which
is \emph{less} than unity: the sibling anticorrelation enforced by
conservation.

The general correlator factorizes over the merger tree.  At depth
$\ell$ the $n$ points occupy cells of the level-$\ell$ grid; grouping
those cells by their common parent $P$ and writing $s_1(P), \ldots,
s_{j_P}(P)$ for the point counts in the $j_P$ occupied children of $P$,
\begin{equation}
  \Big\langle \prod_{i=1}^{n} \rho(\mathbf{x}_i) \Big\rangle
  = \prod_{\ell=1}^{\Lambda} \prod_{P \in \mathcal{P}_\ell}
    M_{\,s_1(P) \cdots s_{j_P}(P)},
  \label{eq:npoint}
\end{equation}
where $\mathcal{P}_\ell$ is the set of occupied parents at depth
$\ell$.  A parent whose single occupied child holds a single point
contributes $\langle m \rangle = 1$, so the product terminates once
every point is isolated.  Each level applies the same linear kernel
(Equation~\ref{eq:sibmom}) to the occupancy configuration: the
transport-operator structure of \citet{greiner98}.

For the two-point function the placement average is closed-form.  With
$\mathbf{s}$ the pair separation in box units, the probability that
both points share one depth-$\ell$ cell is
$Q_\ell(\mathbf{s}) = \prod_{a=1}^{d} (1 - 2^{\ell} |s_a|)_+$, and
\begin{equation}
  1 + \xi(\mathbf{s})
  = \frac{1}{Q_0(\mathbf{s})} \left[
      Q_\Lambda(\mathbf{s})\, M_2^{\Lambda}
      + c_2 \sum_{\ell=0}^{\Lambda-1}
        \big( Q_\ell(\mathbf{s}) - Q_{\ell+1}(\mathbf{s}) \big)
        M_2^{\ell}
    \right],
  \label{eq:xicascade}
\end{equation}
with the isotropic $\xi(r)$ obtained by averaging over orientations of
$\mathbf{s}$.  Since $c_2 < 1$, $\xi$ turns negative at separations
approaching the box scale: conservation builds the integral constraint
into the field itself.  Density-contrast correlators follow from
Equation~(\ref{eq:npoint}) by inclusion--exclusion,
\begin{equation}
  \langle \delta_1 \cdots \delta_n \rangle
  = \sum_{T \subseteq \{1, \ldots, n\}} (-1)^{\,n - |T|}
    \Big\langle \prod_{i \in T} \rho_i \Big\rangle ,
  \label{eq:inclexcl}
\end{equation}
with $\langle \rho \rangle = 1$, and the full, disconnected, and
connected 4pCF from Equations~(\ref{eq:disc}) and (\ref{eq:conn}) with
the $\xi$ of Equation~(\ref{eq:xicascade}).  For three- and four-point
configurations the geometry enters only through the joint
cell-occupancy patterns of Equation~(\ref{eq:npoint}), which we average
over placements and orientations in the estimator's own bin measure
(Section~\ref{sec:solvable}). 

{
\textit{The four-point case.}  For the 4pCF validation
Equation~(\ref{eq:npoint}) takes an explicit form.  Let $\ell_{ab}$
($0 \le \ell_{ab} \le \Lambda$) be the depth of the smallest cell
shared by points $a$ and $b$; the dyadic nesting makes these depths
ultrametric, so they define the merger tree of the four points.  With
$L = \min_{ab} \ell_{ab}$, the four points occupy a single child at
each of the levels $1, \ldots, L$ and first separate at level $L+1$, so
that, writing $\{a,b,c,d\}$ for a relabelling of $\{1,2,3,4\}$ according
to the partition at that level,
\begin{equation}
  \langle \rho_1 \rho_2 \rho_3 \rho_4 \rangle = M_4^{L} \times
  \begin{cases}
    M_{1111} & \{a\},\{b\},\{c\},\{d\}, \\
    M_{211}\, T_{ab}(L) & \{a,b\},\{c\},\{d\}, \\
    M_{22}\, T_{ab}(L)\, T_{cd}(L) & \{a,b\},\{c,d\}, \\
    M_{31}\, M_3^{\,L_3 - L - 1}\, G_{abc} & \{a,b,c\},\{d\},
  \end{cases}
  \label{eq:rho4}
\end{equation}
or $M_4^{\Lambda}$ if all four points share a leaf cell.  Here
$T_{ab}(p) = M_2^{\,\ell_{ab} - p - 1}\, c_2^{[\ell_{ab} < \Lambda]}$ is
the factor of a pair that becomes a separate group at level $p+1$ and
separates at level $\ell_{ab} + 1$ (the $c_2$ is absent if the pair shares
a leaf); $L_3 = \min(\ell_{ab}, \ell_{ac}, \ell_{bc})$ is the depth to
which a triple stays together, and $G_{abc}$ is $M_{111}$ if the triple
then splits into three single points, $M_{21}\, T_{ab}(L_3)$ if it splits
off the pair $\{a,b\}$, and unity if $L_3 = \Lambda$.  The same rules give
$\langle \rho_a \rho_b \rangle = M_2^{\,\ell_{ab}}\,
c_2^{[\ell_{ab} < \Lambda]}$ and
$\langle \rho_a \rho_b \rho_c \rangle = M_3^{\,L_3}\, G_{abc}$, and
Equation~(\ref{eq:inclexcl}) becomes
\begin{equation}
  \langle \delta_1 \delta_2 \delta_3 \delta_4 \rangle
  = \langle \rho_1 \rho_2 \rho_3 \rho_4 \rangle
  - \sum_{\rm 4\ triples} \langle \rho_a \rho_b \rho_c \rangle
  + \sum_{\rm 6\ pairs} \langle \rho_a \rho_b \rangle - 3 ,
  \label{eq:delta4}
\end{equation}
whose average over a configuration's bin cell is the predicted full 4pCF;
subtracting the pairing sum of Equation~(\ref{eq:disc}) with
$\xi_{ab} = \langle \rho_a \rho_b \rangle - 1$ gives the connected one.
With the power sums $S_q = \sum_i w_i^q$, Equation~(\ref{eq:sibmom})
gives for $d = 3$ the sibling moments needed:
$M_q = 2^{3(q-1)} S_q$ ($q = 2, 3, 4$),
$c_2 = 2^6 (1 - S_2)/56$,
$M_{21} = 2^9 (S_2 - S_3)/56$,
$M_{111} = 2^9 (1 - 3S_2 + 2S_3)/336$,
$M_{31} = 2^{12} (S_3 - S_4)/56$,
$M_{22} = 2^{12} (S_2^2 - S_4)/56$,
$M_{211} = 2^{12} (S_2 - S_2^2 - 2S_3 + 2S_4)/336$ and
$M_{1111} = 2^{12} (1 - 6S_2 + 3S_2^2 + 8S_3 - 6S_4)/1680$.
For the weights used here these are
$M_2 = 1.1995$, $M_3 = 1.6746$, $M_4 = 2.6218$, $c_2 = 0.9715$,
$M_{21} = 1.1316$, $M_{111} = 0.9181$, $M_{31} = 1.5392$,
$M_{22} = 1.2697$, $M_{211} = 1.0406$ and $M_{1111} = 0.8446$.
}

\bibliography{gramsci_v2}

\begin{thebibliography}{}
\expandafter\ifx\csname natexlab\endcsname\relax\def\natexlab#1{#1}\fi
\providecommand{\url}[1]{\href{#1}{#1}}
\providecommand{\dodoi}[1]{doi:~\href{http://doi.org/#1}{\nolinkurl{#1}}}
\providecommand{\doeprint}[1]{\href{http://ascl.net/#1}{\nolinkurl{http://ascl.net/#1}}}
\providecommand{\doarXiv}[1]{\href{https://arxiv.org/abs/#1}{\nolinkurl{https://arxiv.org/abs/#1}}}

\bibitem[{A.~G. {Adame} {et~al.}(2025){Adame}, {Aguilar}, {Ahlen}, {Alam},
  {Alexander}, {Alvarez}, {Alves}, {Anand}, {Andrade}, {Armengaud}, {Avila},
  {Aviles}, {Awan}, {Bahr-Kalus}, {Bailey}, {Baltay}, {Bault}, {Behera},
  {BenZvi}, {Bera}, {Beutler}, {Bianchi}, {Blake}, {Blum}, {Brieden},
  {Brodzeller}, {Brooks}, {Buckley-Geer}, {Burtin}, {Calderon}, {Canning},
  {Carnero Rosell}, {Cereskaite}, {Cervantes-Cota}, {Chabanier}, {Chaussidon},
  {Chaves-Montero}, {Chen}, {Chen}, {Claybaugh}, {Cole}, {Cuceu}, {Davis},
  {Dawson}, {de la Macorra}, {de Mattia}, {Deiosso}, {Dey}, {Dey}, {Ding},
  {Doel}, {Edelstein}, {Eftekharzadeh}, {Eisenstein}, {Elliott}, {Fagrelius},
  {Fanning}, {Ferraro}, {Ereza}, {Findlay}, {Flaugher}, {Font-Ribera},
  {Forero-S{\'a}nchez}, {Forero-Romero}, {Frenk}, {Garcia-Quintero},
  {Gazta{\~n}aga}, {Gil-Mar{\'\i}n}, {Gontcho a Gontcho}, {Gonzalez-Morales},
  {Gonzalez-Perez}, {Gordon}, {Green}, {Gruen}, {Gsponer}, {Gutierrez}, {Guy},
  {Hadzhiyska}, {Hahn}, {Hanif}, {Herrera-Alcantar}, {Honscheid}, {Howlett},
  {Huterer}, {Ir{\v{s}}i{\v{c}}}, {Ishak}, {Juneau}, {Kara{\c{c}}ayl{\i}},
  {Kehoe}, {Kent}, {Kirkby}, {Kremin}, {Krolewski}, {Lai}, {Lan}, {Landriau},
  {Lang}, {Lasker}, {Le Goff}, {Le Guillou}, {Leauthaud}, {Levi}, {Li},
  {Linder}, {Lodha}, {Magneville}, {Manera}, {Margala}, {Martini}, {Maus},
  {McDonald}, {Medina-Varela}, {Meisner}, {Mena-Fern{\'a}ndez}, {Miquel},
  {Moon}, {Moore}, {Moustakas}, {Mueller}, {Mu{\~n}oz-Guti{\'e}rrez}, {Myers},
  {Nadathur}, {Napolitano}, {Neveux}, {Newman}, {Nguyen}, {Nie}, {Niz},
  {Noriega}, {Padmanabhan}, {Paillas}, {Palanque-Delabrouille}, {Pan},
  {Penmetsa}, {Percival}, {Pieri}, {Pinon}, {Poppett}, {Porredon}, {Prada},
  {P{\'e}rez-Fern{\'a}ndez}, {P{\'e}rez-R{\`a}fols}, {Rabinowitz}, {Raichoor},
  {Ram{\'\i}rez-P{\'e}rez}, {Ramirez-Solano}, {Rashkovetskyi}, {Ravoux},
  {Rezaie}, {Rich}, {Rocher}, {Rockosi}, {Roe}, {Rosado-Marin}, {Ross},
  {Rossi}, {Ruggeri}, {Ruhlmann-Kleider}, {Samushia}, {Sanchez}, {Saulder},
  {Schlafly}, {Schlegel}, {Schubnell}, {Seo}, {Shafieloo}, {Sharples},
  {Silber}, {Slosar}, {Smith}, {Sprayberry}, {Tan}, {Tarl{\'e}}, {Taylor},
  {Trusov}, {Ure{\~n}a-L{\'o}pez}, {Vaisakh}, {Valcin}, {Valdes},
  {Vargas-Maga{\~n}a}, {Verde}, {Walther}, {Wang}, {Wang}, {Weaver},
  {Weaverdyck}, {Wechsler}, {Weinberg}, {White}, {Yu}, {Yu}, {Yuan},
  {Y{\`e}che}, {Zaborowski}, {Zarrouk}, {Zhang}, {Zhao}, {Zhao}, {Zhou}, \&
  {Zhuang}}]{desi24}
{Adame}, A.~G., {Aguilar}, J., {Ahlen}, S., {et~al.} 2025,
  \bibinfo{title}{{DESI 2024 VI: cosmological constraints from the measurements
  of baryon acoustic oscillations},} \jcap, 2025, 021,
  \dodoi{10.1088/1475-7516/2025/02/021}

\bibitem[{S. {Alam} {et~al.}(2017){Alam}, {Ata}, {Bailey}, {Beutler},
  {Bizyaev}, {Blazek}, {Bolton}, {Brownstein}, {Burden}, {Chuang}, {Comparat},
  {Cuesta}, {Dawson}, {Eisenstein}, {Escoffier}, {Gil-Mar{\'\i}n}, {Grieb},
  {Hand}, {Ho}, {Kinemuchi}, {Kirkby}, {Kitaura}, {Malanushenko},
  {Malanushenko}, {Maraston}, {McBride}, {Nichol}, {Olmstead}, {Oravetz},
  {Padmanabhan}, {Palanque-Delabrouille}, {Pan}, {Pellejero-Ibanez},
  {Percival}, {Petitjean}, {Prada}, {Price-Whelan}, {Reid},
  {Rodr{\'\i}guez-Torres}, {Roe}, {Ross}, {Ross}, {Rossi},
  {Rubi{\~n}o-Mart{\'\i}n}, {Saito}, {Salazar-Albornoz}, {Samushia},
  {S{\'a}nchez}, {Satpathy}, {Schlegel}, {Schneider}, {Sc{\'o}ccola}, {Seo},
  {Sheldon}, {Simmons}, {Slosar}, {Strauss}, {Swanson}, {Thomas}, {Tinker},
  {Tojeiro}, {Maga{\~n}a}, {Vazquez}, {Verde}, {Wake}, {Wang}, {Weinberg},
  {White}, {Wood-Vasey}, {Y{\`e}che}, {Zehavi}, {Zhai}, \& {Zhao}}]{alam17}
{Alam}, S., {Ata}, M., {Bailey}, S., {et~al.} 2017, \bibinfo{title}{{The
  clustering of galaxies in the completed SDSS-III Baryon Oscillation
  Spectroscopic Survey: cosmological analysis of the DR12 galaxy sample},}
  \mnras, 470, 2617, \dodoi{10.1093/mnras/stx721}

\bibitem[{L. {Anderson} {et~al.}(2014){Anderson}, {Aubourg}, {Bailey},
  {Beutler}, {Bhardwaj}, {Blanton}, {Bolton}, {Brinkmann}, {Brownstein},
  {Burden}, {Chuang}, {Cuesta}, {Dawson}, {Eisenstein}, {Escoffier}, {Gunn},
  {Guo}, {Ho}, {Honscheid}, {Howlett}, {Kirkby}, {Lupton}, {Manera},
  {Maraston}, {McBride}, {Mena}, {Montesano}, {Nichol}, {Nuza}, {Olmstead},
  {Padmanabhan}, {Palanque-Delabrouille}, {Parejko}, {Percival}, {Petitjean},
  {Prada}, {Price-Whelan}, {Reid}, {Roe}, {Ross}, {Ross}, {Sabiu}, {Saito},
  {Samushia}, {S{\'a}nchez}, {Schlegel}, {Schneider}, {Scoccola}, {Seo},
  {Skibba}, {Strauss}, {Swanson}, {Thomas}, {Tinker}, {Tojeiro}, {Maga{\~n}a},
  {Verde}, {Wake}, {Weaver}, {Weinberg}, {White}, {Xu}, {Y{\`e}che}, {Zehavi},
  \& {Zhao}}]{anderson14}
{Anderson}, L., {Aubourg}, {\'E}., {Bailey}, S., {et~al.} 2014,
  \bibinfo{title}{{The clustering of galaxies in the SDSS-III Baryon
  Oscillation Spectroscopic Survey: baryon acoustic oscillations in the Data
  Releases 10 and 11 Galaxy samples},} \mnras, 441, 24,
  \dodoi{10.1093/mnras/stu523}

\bibitem[{ {Astropy Collaboration} {et~al.}(2013){Astropy Collaboration},
  {Robitaille}, {Tollerud}, {Greenfield}, {Droettboom}, {Bray}, {Aldcroft},
  {Davis}, {Ginsburg}, {Price-Whelan}, {Kerzendorf}, {Conley}, {Crighton},
  {Barbary}, {Muna}, {Ferguson}, {Grollier}, {Parikh}, {Nair}, {Unther},
  {Deil}, {Woillez}, {Conseil}, {Kramer}, {Turner}, {Singer}, {Fox}, {Weaver},
  {Zabalza}, {Edwards}, {Azalee Bostroem}, {Burke}, {Casey}, {Crawford},
  {Dencheva}, {Ely}, {Jenness}, {Labrie}, {Lim}, {Pierfederici}, {Pontzen},
  {Ptak}, {Refsdal}, {Servillat}, \& {Streicher}}]{astropy13}
{Astropy Collaboration}, {Robitaille}, T.~P., {Tollerud}, E.~J., {et~al.} 2013,
  \bibinfo{title}{{Astropy: A community Python package for astronomy},} \aap,
  558, A33, \dodoi{10.1051/0004-6361/201322068}

\bibitem[{ {Astropy Collaboration} {et~al.}(2018){Astropy Collaboration},
  {Price-Whelan}, {Sip{\H{o}}cz}, {G{\"u}nther}, {Lim}, {Crawford}, {Conseil},
  {Shupe}, {Craig}, {Dencheva}, {Ginsburg}, {VanderPlas}, {Bradley},
  {P{\'e}rez-Su{\'a}rez}, {de Val-Borro}, {Aldcroft}, {Cruz}, {Robitaille},
  {Tollerud}, {Ardelean}, {Babej}, {Bach}, {Bachetti}, {Bakanov}, {Bamford},
  {Barentsen}, {Barmby}, {Baumbach}, {Berry}, {Biscani}, {Boquien}, {Bostroem},
  {Bouma}, {Brammer}, {Bray}, {Breytenbach}, {Buddelmeijer}, {Burke},
  {Calderone}, {Cano Rodr{\'\i}guez}, {Cara}, {Cardoso}, {Cheedella}, {Copin},
  {Corrales}, {Crichton}, {D'Avella}, {Deil}, {Depagne}, {Dietrich}, {Donath},
  {Droettboom}, {Earl}, {Erben}, {Fabbro}, {Ferreira}, {Finethy}, {Fox},
  {Garrison}, {Gibbons}, {Goldstein}, {Gommers}, {Greco}, {Greenfield},
  {Groener}, {Grollier}, {Hagen}, {Hirst}, {Homeier}, {Horton}, {Hosseinzadeh},
  {Hu}, {Hunkeler}, {Ivezi{\'c}}, {Jain}, {Jenness}, {Kanarek}, {Kendrew},
  {Kern}, {Kerzendorf}, {Khvalko}, {King}, {Kirkby}, {Kulkarni}, {Kumar},
  {Lee}, {Lenz}, {Littlefair}, {Ma}, {Macleod}, {Mastropietro}, {McCully},
  {Montagnac}, {Morris}, {Mueller}, {Mumford}, {Muna}, {Murphy}, {Nelson},
  {Nguyen}, {Ninan}, {N{\"o}the}, {Ogaz}, {Oh}, {Parejko}, {Parley}, {Pascual},
  {Patil}, {Patil}, {Plunkett}, {Prochaska}, {Rastogi}, {Reddy Janga},
  {Sabater}, {Sakurikar}, {Seifert}, {Sherbert}, {Sherwood-Taylor}, {Shih},
  {Sick}, {Silbiger}, {Singanamalla}, {Singer}, {Sladen}, {Sooley},
  {Sornarajah}, {Streicher}, {Teuben}, {Thomas}, {Tremblay}, {Turner},
  {Terr{\'o}n}, {van Kerkwijk}, {de la Vega}, {Watkins}, {Weaver}, {Whitmore},
  {Woillez}, {Zabalza}, \& {Astropy Contributors}}]{astropy18}
{Astropy Collaboration}, {Price-Whelan}, A.~M., {Sip{\H{o}}cz}, B.~M., {et~al.}
  2018, \bibinfo{title}{{The Astropy Project: Building an Open-science Project
  and Status of the v2.0 Core Package},} \aj, 156, 123,
  \dodoi{10.3847/1538-3881/aabc4f}

\bibitem[{ {Astropy Collaboration} {et~al.}(2022){Astropy Collaboration},
  {Price-Whelan}, {Lim}, {Earl}, {Starkman}, {Bradley}, {Shupe}, {Patil},
  {Corrales}, {Brasseur}, {N{\"o}the}, {Donath}, {Tollerud}, {Morris},
  {Ginsburg}, {Vaher}, {Weaver}, {Tocknell}, {Jamieson}, {van Kerkwijk},
  {Robitaille}, {Merry}, {Bachetti}, {G{\"u}nther}, {Aldcroft},
  {Alvarado-Montes}, {Archibald}, {B{\'o}di}, {Bapat}, {Barentsen},
  {Baz{\'a}n}, {Biswas}, {Boquien}, {Burke}, {Cara}, {Cara}, {Conroy},
  {Conseil}, {Craig}, {Cross}, {Cruz}, {D'Eugenio}, {Dencheva}, {Devillepoix},
  {Dietrich}, {Eigenbrot}, {Erben}, {Ferreira}, {Foreman-Mackey}, {Fox},
  {Freij}, {Garg}, {Geda}, {Glattly}, {Gondhalekar}, {Gordon}, {Grant},
  {Greenfield}, {Groener}, {Guest}, {Gurovich}, {Handberg}, {Hart},
  {Hatfield-Dodds}, {Homeier}, {Hosseinzadeh}, {Jenness}, {Jones}, {Joseph},
  {Kalmbach}, {Karamehmetoglu}, {Ka{\l}uszy{\'n}ski}, {Kelley}, {Kern},
  {Kerzendorf}, {Koch}, {Kulumani}, {Lee}, {Ly}, {Ma}, {MacBride}, {Maljaars},
  {Muna}, {Murphy}, {Norman}, {O'Steen}, {Oman}, {Pacifici}, {Pascual},
  {Pascual-Granado}, {Patil}, {Perren}, {Pickering}, {Rastogi}, {Roulston},
  {Ryan}, {Rykoff}, {Sabater}, {Sakurikar}, {Salgado}, {Sanghi}, {Saunders},
  {Savchenko}, {Schwardt}, {Seifert-Eckert}, {Shih}, {Jain}, {Shukla}, {Sick},
  {Simpson}, {Singanamalla}, {Singer}, {Singhal}, {Sinha}, {Sip{\H{o}}cz},
  {Spitler}, {Stansby}, {Streicher}, {{\v{S}}umak}, {Swinbank}, {Taranu},
  {Tewary}, {Tremblay}, {de Val-Borro}, {Van Kooten}, {Vasovi{\'c}}, {Verma},
  {de Miranda Cardoso}, {Williams}, {Wilson}, {Winkel}, {Wood-Vasey}, {Xue},
  {Yoachim}, {Zhang}, {Zonca}, \& {Astropy Project Contributors}}]{astropy22}
{Astropy Collaboration}, {Price-Whelan}, A.~M., {Lim}, P.~L., {et~al.} 2022,
  \bibinfo{title}{{The Astropy Project: Sustaining and Growing a
  Community-oriented Open-source Project and the Latest Major Release (v5.0) of
  the Core Package},} \apj, 935, 167, \dodoi{10.3847/1538-4357/ac7c74}

\bibitem[{Z. {Brown} {et~al.}(2022){Brown}, {Mishtaku}, \& {Demina}}]{brown22}
{Brown}, Z., {Mishtaku}, G., \& {Demina}, R. 2022, \bibinfo{title}{{ConKer: An
  algorithm for evaluating correlations of arbitrary order},} \aap, 667, A129,
  \dodoi{10.1051/0004-6361/202141917}

\bibitem[{G. {Cabass} {et~al.}(2023){Cabass}, {Ivanov}, \&
  {Philcox}}]{cabass23}
{Cabass}, G., {Ivanov}, M.~M., \& {Philcox}, O. H.~E. 2023,
  \bibinfo{title}{{Colliders and ghosts: Constraining inflation with the
  parity-odd galaxy four-point function},} \prd, 107, 023523,
  \dodoi{10.1103/PhysRevD.107.023523}

\bibitem[{R.~N. {Cahn} {et~al.}(2023){Cahn}, {Slepian}, \& {Hou}}]{cahn23}
{Cahn}, R.~N., {Slepian}, Z., \& {Hou}, J. 2023, \bibinfo{title}{{Test for
  Cosmological Parity Violation Using the 3D Distribution of Galaxies},} \prl,
  130, 201002, \dodoi{10.1103/PhysRevLett.130.201002}

\bibitem[{C.-H. {Chuang} {et~al.}(2015){Chuang}, {Kitaura}, {Prada}, {Zhao}, \&
  {Yepes}}]{chuang15}
{Chuang}, C.-H., {Kitaura}, F.-S., {Prada}, F., {Zhao}, C., \& {Yepes}, G.
  2015, \bibinfo{title}{{EZmocks: extending the Zel'dovich approximation to
  generate mock galaxy catalogues with accurate clustering statistics},}
  \mnras, 446, 2621, \dodoi{10.1093/mnras/stu2301}

\bibitem[{W.~R. {Coulton} {et~al.}(2024){Coulton}, {Philcox}, \&
  {Villaescusa-Navarro}}]{coulton23}
{Coulton}, W.~R., {Philcox}, O. H.~E., \& {Villaescusa-Navarro}, F. 2024,
  \bibinfo{title}{{Signatures of a parity-violating universe},} \prd, 109,
  023531, \dodoi{10.1103/PhysRevD.109.023531}

\bibitem[{M. {Davis} \& P.~J.~E. {Peebles}(1983){Davis} \& {Peebles}}]{davis83}
{Davis}, M., \& {Peebles}, P.~J.~E. 1983, \bibinfo{title}{{A survey of galaxy
  redshifts. V. The two-point position and velocity correlations.},} \apj, 267,
  465, \dodoi{10.1086/160884}

\bibitem[{ {DESI Collaboration} {et~al.}(2026){DESI Collaboration}, {Abdul
  Karim}, {Adame}, {Aguado}, {Aguilar}, {Ahlen}, {Alam}, {Aldering},
  {Alexander}, {Alfarsy}, {Allen}, {Allende Prieto}, {Alves}, {Anand},
  {Andrade}, {Armengaud}, {Avila}, {Aviles}, {Awan}, {Bailey}, {Baleato
  Lizancos}, {Ballester}, {Bault}, {Bautista}, {Bean}, {Behera}, {BenZvi},
  {Beraldo e Silva}, {Bermejo-Climent}, {Beutler}, {Bianchi}, {Blake}, {Blum},
  {Bolton}, {Bonici}, {Brieden}, {Brodzeller}, {Brooks}, {Buckley-Geer},
  {Burtin}, {Bystr{\"o}m}, {Canning}, {Carnero Rosell}, {Carr}, {Carrilho},
  {Casas}, {Castander}, {Cereskaite}, {Cervantes-Cota}, {Chaussidon},
  {Chaves-Montero}, {Chen}, {Chen}, {Circosta}, {Claybaugh}, {Cole}, {Cooper},
  {Cousinou}, {Cuceu}, {Davis}, {Dawson}, {de Belsunce}, {de la Cruz}, {de la
  Macorra}, {de Mattia}, {Deiosso}, {Della Costa}, {Demina}, {Demirbozan},
  {DeRose}, {Dey}, {Dey}, {Ding}, {Ding}, {Doel}, {Douglass}, {Dowicz},
  {Ebina}, {Edelstein}, {Eisenstein}, {Elbers}, {Emas}, {Escoffier},
  {Fagrelius}, {Fan}, {Fanning}, {Favole}, {Fawcett},
  {Fern{\'a}ndez-Garc{\'\i}a}, {Ferraro}, {Findlay}, {Font-Ribera},
  {Forero-Romero}, {Forero-S{\'a}nchez}, {Frenk}, {G{\"a}nsicke}, {Galbany},
  {Garc{\'\i}a-Bellido}, {Garcia-Quintero}, {Garrison}, {Gazta{\~n}aga},
  {Gil-Mar{\'\i}n}, {Gloudemans}, {Gnedin}, {Gontcho A Gontcho}, {Gonzalez},
  {Gonzalez-Morales}, {Gonzalez-Perez}, {Gordon}, {Graur}, {Green}, {Gruen},
  {Gsponer}, {Guandalin}, {Gutierrez}, {Guy}, {Hahn}, {Han}, {Han}, {He},
  {Herrera-Alcantar}, {Heydenreich}, {Honscheid}, {Hou}, {Howlett}, {Huterer},
  {Ir{\v{s}}i{\v{c}}}, {Ishak}, {Jacques}, {Jiang}, {Jimenez}, {Jing},
  {Joachimi}, {Joudaki}, {Joyce}, {Jullo}, {Juneau}, {Kara{\c{c}}ayl{\i}},
  {Karim}, {Kehoe}, {Kent}, {Khederlarian}, {Kirkby}, {Kisner}, {Kitaura},
  {Kizhuprakkat}, {Kong}, {Koposov}, {Kremin}, {Krolewski}, {Lahav}, {Lai},
  {Lamman}, {Lan}, {Landriau}, {Lang}, {Lange}, {Lasker}, {Le Goff}, {Le
  Guillou}, {Leauthaud}, {Levi}, {Li}, {Li}, {Liu}, {Lodha}, {Lokken}, {Luo},
  {Magneville}, {Manera}, {Manser}, {Margala}, {Martini}, {Maus}, {McCullough},
  {McDonald}, {Medina}, {Medina-Varela}, {Meisner}, {Mena-Fern{\'a}ndez},
  {Menegas}, {Meneses-Rizo}, {Mezcua}, {Miquel}, {Montero-Camacho}, {Moon},
  {Moustakas}, {Mu{\~n}oz-Guti{\'e}rrez}, {Mu noz-Santos}, {Myers}, {Myles},
  {Nadathur}, {Najita}, {Napolitano}, {Newman}, {Nikakhtar}, {Nikutta}, {Niz},
  {Noriega}, \& {Nugent}}]{desidr1}
{DESI Collaboration}, {Abdul Karim}, M., {Adame}, A.~G., {et~al.} 2026,
  \bibinfo{title}{{Data Release 1 of the Dark Energy Spectroscopic
  Instrument},} \aj, 171, 285, \dodoi{10.3847/1538-3881/ae4c43}

\bibitem[{D.~J. {Eisenstein} {et~al.}(2005){Eisenstein}, {Zehavi}, {Hogg},
  {Scoccimarro}, {Blanton}, {Nichol}, {Scranton}, {Seo}, {Tegmark}, {Zheng},
  {Anderson}, {Annis}, {Bahcall}, {Brinkmann}, {Burles}, {Castander},
  {Connolly}, {Csabai}, {Doi}, {Fukugita}, {Frieman}, {Glazebrook}, {Gunn},
  {Hendry}, {Hennessy}, {Ivezi{\'c}}, {Kent}, {Knapp}, {Lin}, {Loh}, {Lupton},
  {Margon}, {McKay}, {Meiksin}, {Munn}, {Pope}, {Richmond}, {Schlegel},
  {Schneider}, {Shimasaku}, {Stoughton}, {Strauss}, {SubbaRao}, {Szalay},
  {Szapudi}, {Tucker}, {Yanny}, \& {York}}]{eisenstein05}
{Eisenstein}, D.~J., {Zehavi}, I., {Hogg}, D.~W., {et~al.} 2005,
  \bibinfo{title}{{Detection of the Baryon Acoustic Peak in the Large-Scale
  Correlation Function of SDSS Luminous Red Galaxies},} \apj, 633, 560,
  \dodoi{10.1086/466512}

\bibitem[{J.~A. {Frieman} \& E. {Gaztanaga}(1994){Frieman} \&
  {Gaztanaga}}]{frieman94}
{Frieman}, J.~A., \& {Gaztanaga}, E. 1994, \bibinfo{title}{{The Three-Point
  Function as a Probe of Models for Large-Scale Structure},} \apj, 425, 392,
  \dodoi{10.1086/173995}

\bibitem[{J.~N. {Fry} \& E. {Gaztanaga}(1993){Fry} \& {Gaztanaga}}]{fry93}
{Fry}, J.~N., \& {Gaztanaga}, E. 1993, \bibinfo{title}{{Biasing and
  Hierarchical Statistics in Large-Scale Structure},} \apj, 413, 447,
  \dodoi{10.1086/173015}

\bibitem[{E. {Gazta{\~n}aga} {et~al.}(2009){Gazta{\~n}aga}, {Cabr{\'e}},
  {Castander}, {Crocce}, \& {Fosalba}}]{gaztanaga09}
{Gazta{\~n}aga}, E., {Cabr{\'e}}, A., {Castander}, F., {Crocce}, M., \&
  {Fosalba}, P. 2009, \bibinfo{title}{{Clustering of luminous red galaxies -
  III. Baryon acoustic peak in the three-point correlation},} \mnras, 399, 801,
  \dodoi{10.1111/j.1365-2966.2009.15313.x}

\bibitem[{E. {Gazta{\~n}aga} \& R. {Scoccimarro}(2005){Gazta{\~n}aga} \&
  {Scoccimarro}}]{gaztanaga05}
{Gazta{\~n}aga}, E., \& {Scoccimarro}, R. 2005, \bibinfo{title}{{The
  three-point function in large-scale structure: redshift distortions and
  galaxy bias},} \mnras, 361, 824, \dodoi{10.1111/j.1365-2966.2005.09234.x}

\bibitem[{H. {Gil-Mar{\'\i}n} {et~al.}(2017){Gil-Mar{\'\i}n}, {Percival},
  {Verde}, {Brownstein}, {Chuang}, {Kitaura}, {Rodr{\'\i}guez-Torres}, \&
  {Olmstead}}]{gilmarin17}
{Gil-Mar{\'\i}n}, H., {Percival}, W.~J., {Verde}, L., {et~al.} 2017,
  \bibinfo{title}{{The clustering of galaxies in the SDSS-III Baryon
  Oscillation Spectroscopic Survey: RSD measurement from the power spectrum and
  bispectrum of the DR12 BOSS galaxies},} \mnras, 465, 1757,
  \dodoi{10.1093/mnras/stw2679}

\bibitem[{M. {Greiner} {et~al.}(1998){Greiner}, {Eggers}, \&
  {Lipa}}]{greiner98}
{Greiner}, M., {Eggers}, H.~C., \& {Lipa}, P. 1998, \bibinfo{title}{Analytic
  multivariate generating function for random multiplicative cascade
  processes,} Phys. Rev. Lett., 80, 5333

\bibitem[{E.~J. {Groth} \& P.~J.~E. {Peebles}(1977){Groth} \&
  {Peebles}}]{groth77}
{Groth}, E.~J., \& {Peebles}, P.~J.~E. 1977, \bibinfo{title}{{Statistical
  analysis of catalogs of extragalactic objects. VII. Two- and three-point
  correlation functions for the high-resolution Shane-Wirtanen catalog of
  galaxies.},} \apj, 217, 385, \dodoi{10.1086/155588}

\bibitem[{H. {Guo} {et~al.}(2015){Guo}, {Zheng}, {Jing}, {Zehavi}, {Li},
  {Weinberg}, {Skibba}, {Nichol}, {Rossi}, {Sabiu}, {Schneider}, \&
  {McBride}}]{guo15}
{Guo}, H., {Zheng}, Z., {Jing}, Y.~P., {et~al.} 2015,
  \bibinfo{title}{{Modelling the redshift-space three-point correlation
  function in SDSS-III.},} \mnras, 449, L95, \dodoi{10.1093/mnrasl/slv020}

\bibitem[{A.~J.~S. {Hamilton}(1993){Hamilton}}]{hamilton93}
{Hamilton}, A.~J.~S. 1993, \bibinfo{title}{{Toward Better Ways to Measure the
  Galaxy Correlation Function},} \apj, 417, 19, \dodoi{10.1086/173288}

\bibitem[{C.~R. {Harris} {et~al.}(2020){Harris}, {Millman}, {van der Walt},
  {Gommers}, {Virtanen}, {Cournapeau}, {Wieser}, {Taylor}, {Berg}, {Smith},
  {Kern}, {Picus}, {Hoyer}, {van Kerkwijk}, {Brett}, {Haldane}, {del R{\'\i}o},
  {Wiebe}, {Peterson}, {G{\'e}rard-Marchant}, {Sheppard}, {Reddy}, {Weckesser},
  {Abbasi}, {Gohlke}, \& {Oliphant}}]{harris20}
{Harris}, C.~R., {Millman}, K.~J., {van der Walt}, S.~J., {et~al.} 2020,
  \bibinfo{title}{{Array programming with NumPy},} \nat, 585, 357,
  \dodoi{10.1038/s41586-020-2649-2}

\bibitem[{J. {Hou} {et~al.}(2023){Hou}, {Slepian}, \& {Cahn}}]{hou23}
{Hou}, J., {Slepian}, Z., \& {Cahn}, R.~N. 2023, \bibinfo{title}{{Measurement
  of parity-odd modes in the large-scale 4-point correlation function of Sloan
  Digital Sky Survey Baryon Oscillation Spectroscopic Survey twelfth data
  release CMASS and LOWZ galaxies},} \mnras, 522, 5701,
  \dodoi{10.1093/mnras/stad1062}

\bibitem[{J. {Hou} {et~al.}(2026){Hou}, {Cahn}, {Aguilar}, {Ahlen}, {Bianchi},
  {Brooks}, {Cahn}, {Claybaugh}, {Doel}, {Forero-Romero}, {Gazta{\~n}aga}, {Le
  Guillou}, {Gutierrez}, {Howlett}, {Huterer}, {Ishak}, {Joyce}, {Kremin},
  {Lahav}, {Lamman}, {Landriau}, {de la Macorra}, {Miquel}, {Nadathur}, {Niz},
  {Percival}, {Prada}, {P{\'e}rez-R{\`a}fols}, {Rossi}, {Sanchez}, {Schlegel},
  {Schubnell}, {Seo}, {Silber}, {Sprayberry}, {Tarl{\'e}}, {Weaver}, \&
  {Zou}}]{hou25}
{Hou}, J., {Cahn}, R.~N., {Aguilar}, J., {et~al.} 2026,
  \bibinfo{title}{{Parity-odd four-point correlation function from the DESI
  data release 1 luminous red galaxy sample},} \prd, 113, 103502,
  \dodoi{10.1103/2dmd-hyt1}

\bibitem[{J.~D. {Hunter}(2007){Hunter}}]{hunter07}
{Hunter}, J.~D. 2007, \bibinfo{title}{{Matplotlib: A 2D Graphics Environment},}
  Computing in Science and Engineering, 9, 90, \dodoi{10.1109/MCSE.2007.55}

\bibitem[{M. {Jarvis} {et~al.}(2004){Jarvis}, {Bernstein}, \&
  {Jain}}]{jarvis04}
{Jarvis}, M., {Bernstein}, G., \& {Jain}, B. 2004, \bibinfo{title}{The skewness
  of the aperture mass statistic,} MNRAS, 352, 338

\bibitem[{M.~B. {Kennel}(2004){Kennel}}]{kennel04}
{Kennel}, M.~B. 2004, \bibinfo{title}{{KDTREE 2: Fortran 95 and C++ software to
  efficiently search for near neighbors in a multi-dimensional Euclidean
  space},} arXiv e-prints, physics/0408067,
  \dodoi{10.48550/arXiv.physics/0408067}

\bibitem[{S. {Kim} {et~al.}(2026){Kim}, {Sabiu}, \& {Park}}]{kim26}
{Kim}, S., {Sabiu}, C.~G., \& {Park}, I. 2026, \bibinfo{title}{{Climbing the
  N-point ladder. Part I. Information in the higher-order configuration-space
  clustering of dark matter halos},} \jcap, 2026, 139,
  \dodoi{10.1088/1475-7516/2026/09/139}

\bibitem[{S.~D. {Landy} \& A.~S. {Szalay}(1993){Landy} \& {Szalay}}]{landy93}
{Landy}, S.~D., \& {Szalay}, A.~S. 1993, \bibinfo{title}{{Bias and Variance of
  Angular Correlation Functions},} \apj, 412, 64, \dodoi{10.1086/172900}

\bibitem[{B.~B. {Mandelbrot}(1974){Mandelbrot}}]{mandelbrot74}
{Mandelbrot}, B.~B. 1974, \bibinfo{title}{Intermittent turbulence in
  self-similar cascades: divergence of high moments and dimension of the
  carrier,} Journal of Fluid Mechanics, 62, 331

\bibitem[{V.~J. {Mart{\'\i}nez} {et~al.}(1990){Mart{\'\i}nez}, {Jones},
  {Dom{\'\i}nguez-Tenreiro}, \& {van de Weygaert}}]{martinez90}
{Mart{\'\i}nez}, V.~J., {Jones}, B.~J.~T., {Dom{\'\i}nguez-Tenreiro}, R., \&
  {van de Weygaert}, R. 1990, \bibinfo{title}{Clustering Paradigms and
  Multifractal Measures,} \apj, 357, 50, \dodoi{10.1086/168890}

\bibitem[{C. {Meneveau} \& K.~R. {Sreenivasan}(1987){Meneveau} \&
  {Sreenivasan}}]{meneveau87}
{Meneveau}, C., \& {Sreenivasan}, K.~R. 1987, \bibinfo{title}{Simple
  multifractal cascade model for fully developed turbulence,} Phys. Rev. Lett.,
  59, 1424

\bibitem[{P.~J.~E. {Peebles} \& M.~G. {Hauser}(1974){Peebles} \&
  {Hauser}}]{peebles75}
{Peebles}, P.~J.~E., \& {Hauser}, M.~G. 1974, \bibinfo{title}{{Statistical
  Analysis of Catalogs of Extragalactic Objects. III. The Shane-Wirtanen and
  Zwicky Catalogs},} \apjs, 28, 19, \dodoi{10.1086/190308}

\bibitem[{O.~H.~E. {Philcox}(2022){Philcox}}]{philcox22parity}
{Philcox}, O. H.~E. 2022, \bibinfo{title}{{Probing parity violation with the
  four-point correlation function of BOSS galaxies},} \prd, 106, 063501,
  \dodoi{10.1103/PhysRevD.106.063501}

\bibitem[{O.~H.~E. {Philcox} {et~al.}(2022){Philcox}, {Slepian}, {Hou},
  {Warner}, {Cahn}, \& {Eisenstein}}]{philcox22}
{Philcox}, O. H.~E., {Slepian}, Z., {Hou}, J., {et~al.} 2022,
  \bibinfo{title}{{ENCORE: an O (N$_{g}$$^{2}$) estimator for galaxy N-point
  correlation functions},} \mnras, 509, 2457, \dodoi{10.1093/mnras/stab3025}

\bibitem[{C. {Sabiu} {et~al.}(2007){Sabiu}, {Teodoro}, \& {Hendry}}]{sabiu07}
{Sabiu}, C., {Teodoro}, L., \& {Hendry}, M. 2007, \bibinfo{title}{Resolving the
  universe with multifractals,} Il Nuovo Cimento B, 122, 1289

\bibitem[{C.~G. {Sabiu}(2026){Sabiu}}]{sabiu26code}
{Sabiu}, C.~G. 2026, csabiu/Gramsci: GRAMSCI v2.5.0, v2.5.0 Zenodo,
  \dodoi{10.5281/zenodo.22091542}

\bibitem[{C.~G. {Sabiu} {et~al.}(2019){Sabiu}, {Hoyle}, {Kim}, \&
  {Li}}]{sabiu19}
{Sabiu}, C.~G., {Hoyle}, B., {Kim}, J., \& {Li}, X.-D. 2019,
  \bibinfo{title}{{Graph Database Solution for Higher-order Spatial Statistics
  in the Era of Big Data},} \apjs, 242, 29, \dodoi{10.3847/1538-4365/ab22b5}

\bibitem[{E. {Sefusatti} {et~al.}(2006){Sefusatti}, {Crocce}, {Pueblas}, \&
  {Scoccimarro}}]{sefusatti06}
{Sefusatti}, E., {Crocce}, M., {Pueblas}, S., \& {Scoccimarro}, R. 2006,
  \bibinfo{title}{{Cosmology and the bispectrum},} \prd, 74, 023522,
  \dodoi{10.1103/PhysRevD.74.023522}

\bibitem[{E. {Sefusatti} \& E. {Komatsu}(2007){Sefusatti} \&
  {Komatsu}}]{sefusatti07}
{Sefusatti}, E., \& {Komatsu}, E. 2007, \bibinfo{title}{{Bispectrum of galaxies
  from high-redshift galaxy surveys: Primordial non-Gaussianity and nonlinear
  galaxy bias},} \prd, 76, 083004, \dodoi{10.1103/PhysRevD.76.083004}

\bibitem[{M. {Sinha} \& L.~H. {Garrison}(2020){Sinha} \& {Garrison}}]{sinha20}
{Sinha}, M., \& {Garrison}, L.~H. 2020, \bibinfo{title}{{CORRFUNC - a suite of
  blazing fast correlation functions on the CPU},} \mnras, 491, 3022,
  \dodoi{10.1093/mnras/stz3157}

\bibitem[{Z. {Slepian} \& D.~J. {Eisenstein}(2015){Slepian} \&
  {Eisenstein}}]{slepian15}
{Slepian}, Z., \& {Eisenstein}, D.~J. 2015, \bibinfo{title}{{Computing the
  three-point correlation function of galaxies in O(N\^2) time},} \mnras, 454,
  4142, \dodoi{10.1093/mnras/stv2119}

\bibitem[{Z. {Slepian} \& D.~J. {Eisenstein}(2016){Slepian} \&
  {Eisenstein}}]{slepian16}
{Slepian}, Z., \& {Eisenstein}, D.~J. 2016, \bibinfo{title}{{Accelerating the
  two-point and three-point galaxy correlation functions using Fourier
  transforms},} \mnras, 455, L31, \dodoi{10.1093/mnrasl/slv133}

\bibitem[{Z. {Slepian} {et~al.}(2017{\natexlab{a}}){Slepian}, {Eisenstein},
  {Brownstein}, {Chuang}, {Gil-Mar{\'\i}n}, {Ho}, {Kitaura}, {Percival},
  {Ross}, {Rossi}, {Seo}, {Slosar}, \& {Vargas-Maga{\~n}a}}]{slepian17}
{Slepian}, Z., {Eisenstein}, D.~J., {Brownstein}, J.~R., {et~al.}
  2017{\natexlab{a}}, \bibinfo{title}{{Detection of baryon acoustic oscillation
  features in the large-scale three-point correlation function of SDSS BOSS
  DR12 CMASS galaxies},} \mnras, 469, 1738, \dodoi{10.1093/mnras/stx488}

\bibitem[{Z. {Slepian} {et~al.}(2017{\natexlab{b}}){Slepian}, {Eisenstein},
  {Brownstein}, {Chuang}, {Gil-Mar{\'\i}n}, {Ho}, {Kitaura}, {Percival},
  {Ross}, {Rossi}, {Seo}, {Slosar}, \& {Vargas-Maga{\~n}a}}]{slepian17a}
{Slepian}, Z., {Eisenstein}, D.~J., {Brownstein}, J.~R., {et~al.}
  2017{\natexlab{b}}, \bibinfo{title}{{Detection of baryon acoustic oscillation
  features in the large-scale three-point correlation function of SDSS BOSS
  DR12 CMASS galaxies},} \mnras, 469, 1738, \dodoi{10.1093/mnras/stx488}

\bibitem[{Z. {Slepian} {et~al.}(2025){Slepian}, {Krolewski}, {Greco}, {May},
  {Ortola Leonard}, {Kamalinejad}, {Chellino}, {Reinhard}, {Fernandez},
  {Prada}, {Ahlen}, {Bianchi}, {Brooks}, {Claybaugh}, {de la Macorra}, {de
  Mattia}, {Dey}, {Doel}, {Gaztanaga}, {Gutierrez}, {Honscheid}, {Huterer},
  {Joyce}, {Kehoe}, {Kirkby}, {Kisner}, {Landriau}, {Le Guillou}, {Manera},
  {Meisner}, {Miquel}, {Nadathur}, {Percival}, {Ross}, {Sanchez}, {Schlegel},
  {Schubnell}, {Seo}, {Silber}, {Sprayberry}, \& {Tarle}}]{slepian25}
{Slepian}, Z., {Krolewski}, A., {Greco}, A., {et~al.} 2025,
  \bibinfo{title}{{Measurement of Parity-Violating Modes of the Dark Energy
  Spectroscopic Instrument (DESI) Year 1 Luminous Red Galaxies' 4-Point
  Correlation Function},} arXiv e-prints, arXiv:2508.09133,
  \dodoi{10.48550/arXiv.2508.09133}

\bibitem[{I. {Szapudi} \& A.~S. {Szalay}(1998){Szapudi} \&
  {Szalay}}]{szapudi98}
{Szapudi}, I., \& {Szalay}, A.~S. 1998, \bibinfo{title}{{A New Class of
  Estimators for the N-Point Correlations},} \apjl, 494, L41,
  \dodoi{10.1086/311146}

\bibitem[{H. {Totsuji} \& T. {Kihara}(1969){Totsuji} \& {Kihara}}]{totsuji69}
{Totsuji}, H., \& {Kihara}, T. 1969, \bibinfo{title}{{The Correlation Function
  for the Distribution of Galaxies},} \pasj, 21, 221,
  \dodoi{10.1093/pasj/21.3.221}

\bibitem[{L. {Verde} {et~al.}(2002){Verde}, {Heavens}, {Percival}, {Matarrese},
  {Baugh}, {Bland-Hawthorn}, {Bridges}, {Cannon}, {Cole}, {Colless}, {Collins},
  {Couch}, {Dalton}, {De Propris}, {Driver}, {Efstathiou}, {Ellis}, {Frenk},
  {Glazebrook}, {Jackson}, {Lahav}, {Lewis}, {Lumsden}, {Maddox}, {Madgwick},
  {Norberg}, {Peacock}, {Peterson}, {Sutherland}, \& {Taylor}}]{verde02}
{Verde}, L., {Heavens}, A.~F., {Percival}, W.~J., {et~al.} 2002,
  \bibinfo{title}{{The 2dF Galaxy Redshift Survey: the bias of galaxies and the
  density of the Universe},} \mnras, 335, 432,
  \dodoi{10.1046/j.1365-8711.2002.05620.x}

\bibitem[{F. {Villaescusa-Navarro} {et~al.}(2020){Villaescusa-Navarro}, {Hahn},
  {Massara}, {Banerjee}, {Delgado}, {Ramanah}, {Charnock}, {Giusarma}, {Li},
  {Allys}, {Brochard}, {Uhlemann}, {Chiang}, {He}, {Pisani}, {Obuljen}, {Feng},
  {Castorina}, {Contardo}, {Kreisch}, {Nicola}, {Alsing}, {Scoccimarro},
  {Verde}, {Viel}, {Ho}, {Mallat}, {Wandelt}, \& {Spergel}}]{villaescusa20}
{Villaescusa-Navarro}, F., {Hahn}, C., {Massara}, E., {et~al.} 2020,
  \bibinfo{title}{{The Quijote Simulations},} \apjs, 250, 2,
  \dodoi{10.3847/1538-4365/ab9d82}

\bibitem[{P. {Virtanen} {et~al.}(2020){Virtanen}, {Gommers}, {Oliphant},
  {Haberland}, {Reddy}, {Cournapeau}, {Burovski}, {Peterson}, {Weckesser},
  {Bright}, {van der Walt}, {Brett}, {Wilson}, {Millman}, {Mayorov}, {Nelson},
  {Jones}, {Kern}, {Larson}, {Carey}, {Polat}, {Feng}, {Moore}, {VanderPlas},
  {Laxalde}, {Perktold}, {Cimrman}, {Henriksen}, {Quintero}, {Harris},
  {Archibald}, {Ribeiro}, {Pedregosa}, {van Mulbregt}, \& {SciPy 1. 0
  Contributors}}]{virtanen20}
{Virtanen}, P., {Gommers}, R., {Oliphant}, T.~E., {et~al.} 2020,
  \bibinfo{title}{{SciPy 1.0: fundamental algorithms for scientific computing
  in Python},} Nature Medicine, 17, 261, \dodoi{10.1038/s41592-019-0686-2}

\bibitem[{I. {Zehavi} {et~al.}(2011){Zehavi}, {Zheng}, {Weinberg}, {Blanton},
  {Bahcall}, {Berlind}, {Brinkmann}, {Frieman}, {Gunn}, {Lupton}, {Nichol},
  {Percival}, {Schneider}, {Skibba}, {Strauss}, {Tegmark}, \&
  {York}}]{zehavi11}
{Zehavi}, I., {Zheng}, Z., {Weinberg}, D.~H., {et~al.} 2011,
  \bibinfo{title}{{Galaxy Clustering in the Completed SDSS Redshift Survey: The
  Dependence on Color and Luminosity},} \apj, 736, 59,
  \dodoi{10.1088/0004-637X/736/1/59}

\end{thebibliography}
\bibliographystyle{aasjournalv7}

\end{document}